\documentclass[twocolumn]{autart}    % Enable this line and disable the 
\usepackage{graphicx}          % Include this line if your 
\usepackage{subcaption}
\usepackage{amssymb,amsmath}
\usepackage{color}
\usepackage{tikz,pgfplots}
\usepackage{hyperref}

\newtheorem{remark}{Remark}

\usepackage{textcomp}
\usetikzlibrary{shapes,arrows,positioning}
\usepackage{algorithm}
\usepackage[noend]{algpseudocode}

\newcommand{\mbb}[1]{\mathbb #1}

\usepackage{xspace}
\newcommand{\SCC}{\textsc{SC-CO$_2$}\xspace} 
\newcommand{\DICE}{\textsc{DICE}\xspace} 
\newcommand{\MPCDICE}{\textsc{MPC-DICE}\xspace}
\newcommand{\PAGE}{\textsc{PAGE}\xspace} 
\newcommand{\FUND}{\textsc{FUND}\xspace}

\usepackage{color}
\newcommand\CMK[1]{{\color{black} #1}}

\newcommand\TF[1]{{\color{black} #1}}

\begin{document}

\begin{frontmatter}

\title{Feedback, Dynamics, and Optimal Control in \\Climate Economics\thanksref{footnoteinfo}} % Title, preferably not more 

\thanks[footnoteinfo]{This paper was not presented at any IFAC 
meeting.} 
%Corresponding author M.~T.~Cicero. Tel. +XXXIX-VI-mmmxxi. 
%Fax +XXXIX-VI-mmmxxv.}

\author[Newcastle]{Christopher M. Kellett}\ead{chris.kellett@newcastle.edu.au},    % Add the 
\author[Newcastle]{Steven R. Weller}\ead{steven.weller@newcastle.edu.au},
\author[Karlsruhe]{Timm Faulwasser}\ead{timm.faulwasser@ieee.org},  % (ead) as shown
\author[Bayreuth]{Lars Gr\"une}\ead{lars-gruene@uni-bayreuth.de},               % e-mail address 
\author[NewSchool]{Willi Semmler}\ead{semmlerw@newschool.edu}

\address[Newcastle]{School of Electrical Engineering and Computing, University of Newcastle, Callaghan, New South Wales 2308, Australia}  % Please supply                                              
\address[Karlsruhe]{Institute for Automation and Applied Informatics, Karlsruhe Institute of Technology, 76344 Eggenstein-Leopoldshafen, Germany}        % here.
\address[Bayreuth]{Mathematisches Institut, Universit\"at Bayreuth, 95440 Bayreuth, Germany}             % full addresses
\address[NewSchool]{New School for Social Research, New York, United States of America, and University of Bielefeld, Germany, and International Institute for Applied Systems Analysis,  Laxenburg, Austria}


\begin{abstract}                          % Abstract of not more than 200 words.
	For his work in the economics of climate change, Professor William Nordhaus was
	a co-recipient of the 2018 Nobel Memorial Prize for Economic Sciences.  A core
	component of the work undertaken by Nordhaus is the Dynamic
	Integrated model of Climate and Economy, known as the DICE model.  The DICE model
	is a discrete-time model with two control inputs and is primarily used in conjunction
	with a particular optimal control problem in order to estimate optimal pathways 
	for reducing greenhouse gas emissions.  In this paper, \CMK{we provide a tutorial introduction
	to the DICE model and we indicate challenges and open problems of potential interest for
	the systems and control community.}
\end{abstract}

\end{frontmatter}

\section{Introduction}

%Our papers: 
\nocite{Faulwasser-FAIR-DICE-IAMES2018,Faulwasser-MPC-DICE-IAMES2018}
\nocite{HafeezAuCC2015,HafeezAuCC2016,HafeezIFAC2017}
\nocite{Kellett-Humboldt2018}
\nocite{WellerAAEC2015,WellerCCA2015,WellerCDC2015,WellerIFAC2014}

In the absence of deep and sustained reductions in greenhouse gas emissions, the overwhelming scientific consensus points to global warming of several degrees Celsius by 2100. Warming of this magnitude poses profound risks to both human society and natural ecosystems \cite{AR5SPM}. In response to these risks, in late 2015 at the United Nations Climate Change Conference governments around the world committed to urgent reductions in human-caused emissions of greenhouse gases, most notably carbon dioxide (CO$_2$), in order to limit the increase in global average temperature to well below $2~^\circ$C relative to pre-industrial levels.

With global average warming of $1~^\circ$C having already been realized, constraining temperature increases below agreed target levels will require careful control of future emissions, with the Intergovernmental Panel on Climate Change (IPCC) special report on Global Warming of $1.5~^\circ$C \cite{IPCC-1_5-Report} indicating that remaining below $1.5^\circ$C will require net-zero CO$_2$ emissions by about 2050.
Complicating this task are large uncertainties regarding the speed and extent of warming in response to elevated atmospheric CO$_2$ concentrations, coupled with the need for a policy response that balances reduced economic consumption today with avoided (and discounted) economic damages of an uncertain magnitude in the future.

To quantify the damages from anthropogenic emissions of heat-trapping greenhouse gases, specifically CO$_2$, economists model the dynamics of climate--economy interactions using \emph{Integrated Assessment Models (IAMs)}, which incorporate mathematical models of phenomena from both economics and geophysical science.  Possibly the first IAM in the area of climate economics was proposed
by William Nordhaus in \cite{Nordhaus75}.  Subsequently, Nordhaus proposed the Dynamic Integrated model of Climate and Economy (DICE) in \cite{Nordhaus-1992}, with regular refinements and parameter updates, such as \cite{Nordhaus-2008,Nordhaus-PNAS2010,Nordhaus2014,Nordhaus17-PNAS,DICEManual}.  Largely for this body of work, Nordhaus was awarded
the 2018 Nobel Memorial Prize in Economic Sciences.

A central role for IAMs is to estimate the \emph{Social Cost of Carbon Dioxide (SC-CO$_2$)}, defined as the dollar value of the economic damage caused by a one metric tonne increase in CO$_2$ emissions to the atmosphere. 
The \SCC  is used by governments, companies, and international finance organizations as a
key quantity in all aspects of climate change mitigation and adaptation, including cost-benefit analyses, emissions trading schemes, carbon taxes, quantification of energy subsidies, and modelling the impact of climate change on financial assets, known as the value at risk \cite{Economist2015}.  The \SCC therefore underpins trillions of dollars worth of investment decisions \cite{Hope2015}.

A commonly used derivation for the \SCC  solves an open-loop optimal control problem to determine economically optimal CO$_2$ emissions pathways. The open-loop use of IAMs for decision-making, however, disregards crucial uncertainties in both geophysical and economic models. As a consequence, current \SCC  estimates range from US\$11 per tonne of CO$_2$ to US\$63 per tonne of CO$_2$ or higher, and hence the \SCC  fails to reflect the true economic %financial 
risk posed by CO$_2$ emissions, seriously compromising the accuracy of the \SCC  as a price on carbon for the  purposes of climate change mitigation and adaptation \cite{Howarth2014,Otto2015,Pizer2014}.

The IAM community presently pre-dominantly employs simplistic Monte Carlo-based methods to emulate the impact of parametric uncertainty on the \SCC \cite{InteragencyWorkingGroup}, whilst recognizing that such an approach can lead to contradictory policy advice \cite{Crost2013}. On the other hand, known resolutions to this major deficiency are computationally intractable (e.g., stochastic dynamic programming  \cite{Crost2013}) and do not easily accommodate enhanced geophysical models.  At a time when governments, financial bodies \cite{IMF2016}, business \cite{WorldBank2014}, and even emissions-intensive industries \cite{AGLBoss,BP2017} are demanding a price on carbon, it is imperative that the shortcomings in quantifying uncertainty in \SCC estimates be rectified.  Indeed, this is considered to be a problem of the utmost importance \cite{NationalAcademies2016,NatAcadsUpdatingSCC,Hope2015,SternNature2016}.

Our contribution in this paper is three-fold.  First, we provide a complete and replicable specification of the \DICE model, with accompanying code available for download at \cite{MPC-DICE-Code}.  Second, we summarize some of our recent work and update the numerical results to account for updated parameters released by Nordhaus in 2016 \cite{Nordhaus17-PNAS}.  Third, we indicate %some 
\CMK{challenges and
open problems of potential interest for the systems and control community.}

The paper is organized as follows.  Section~\ref{sec:DICE} provides a \CMK{tutorial 
description of the DICE model and of its usage in the context of computing the 
Social Cost of Carbon Dioxide}.  Section~\ref{sec:RHC} describes the benefits of 
receding horizon control for the \DICE model both as a numerical solution technique and
as a way to investigate the impact of parametric uncertainty.  Section~\ref{sec:Constraints} 
considers the impact of placing constraints on the atmospheric temperature and mitigation rate 
constraints.  Section~\ref{sec:Opportunities} \CMK{indicates potential challenges and opportunities
of particular relevance for the systems and control community.
Section~\ref{sec:Summary}
provides some brief concluding remarks.}

\section{The \DICE Model and Methodology}
\label{sec:DICE}

There are three dominant IAMs used for the calculation of the Social Cost of Carbon Dioxide \cite{InteragencyWorkingGroup,Bonen14}: the previously mentioned DICE \cite{DICEManual,Nordhaus2014}, Policy Analysis of the Greenhouse Effect (PAGE) \cite{Hope2013}, and Climate Framework for Uncertainty, Negotiation, and Distribution (FUND) \cite{Antoff2013}.  As we will describe below, the \DICE model and methodology consists of an optimal control problem for a discrete time nonlinear system.  A brief description of \PAGE and \FUND is provided in Section~\ref{sec:PAGEFUND}.

\subsection{DICE Dynamics}
\label{sec:dynamics}
 It should be noted that there exist different open-source implementations of DICE.  While Nordhaus maintains an open-source GAMS implementation \cite{DICECode},\footnote{We refer to \url{https://www.gams.com} for details on GAMS.} a subset of the authors of this paper have recently published open-source \DICE code that runs in Matlab \cite{Matlab-DICE} and \cite{MPC-DICE-Code} (see also \cite{Faulwasser-MPC-DICE-IAMES2018}).

It is important to note at the outset that there is not a definitive statement of \textit{the} \DICE model.
Rather, there are two primary sources in the form of a user's manual \cite{DICEManual} (updated in 
\cite{Nordhaus-NBER2017}) and the available code itself
(both the manual and the code are available at \cite{DICECode}). Additional explanations and, occasionally, 
equations can be found in various other sources including \cite{Nordhaus2014,Nordhaus17-PNAS}.  However, 
these sources are not consistent with each other and, in fact, the specification in
\cite{DICEManual} is incomplete.  Furthermore, there are some minor inconsistencies between 
text and equations in \cite{DICEManual}.  For completeness, and with the aim of presenting
the \DICE model and methodology in a way that can be independently implemented, we
necessarily deviate from \cite{DICEManual} and \cite{DICECode}.  
 However, 
the subsequent
impact on the numerical results {\it when using the default parameters} \CMK{(included in the Appendices)}
is not significant.

One further note before proceeding to the model description: while the most recent version of the model is 
DICE2016 (as used in, for example, \cite{Nordhaus17-PNAS}),
the previous version of the model, DICE2013, has been widely used in the literature.  Importantly, the move
from DICE2013 to DICE2016 involves essentially no structural changes.  Rather, the model update involves
updates on initial conditions and most of the model parameters.  For ease of reference, we provide initial
conditions and model parameters for both DICE2013 and DICE2016 in the appendices.
The Matlab DICE code \cite{MPC-DICE-Code,Matlab-DICE} implements both parameter sets.

\tikzset{%
  block/.style    = {draw, thick, rectangle, minimum height = 4em,
    minimum width = 7em},
  sum/.style      = {draw, circle, node distance = 2.5cm}, % Adder
  input/.style    = {coordinate}, % Input
  output/.style   = {coordinate} % Output
}
% Defining string as labels of certain blocks.
\newcommand{\suma}{\Large$+$}
\tikzstyle{output} = [coordinate]

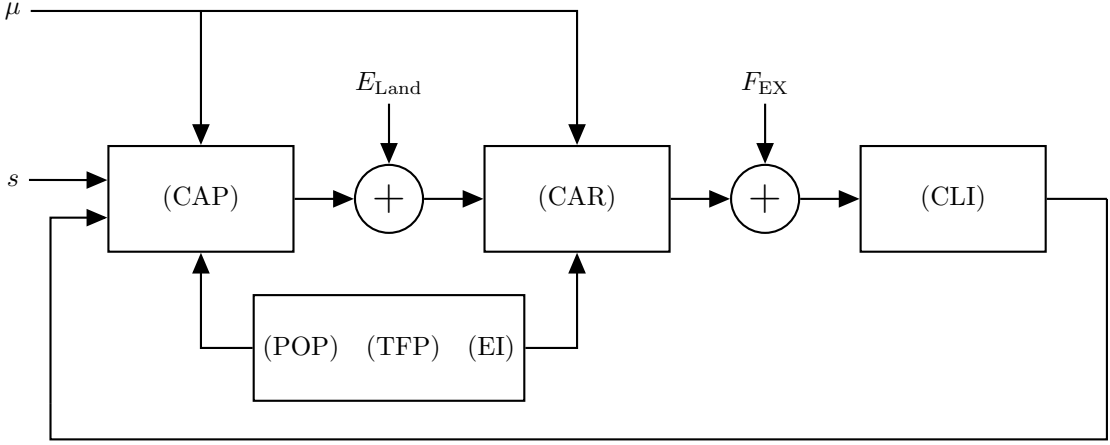
\begin{figure*}[!t]
\begin{center}
\begin{tikzpicture}[auto, thick, node distance=2.5cm, >=triangle 45]
\draw
	% Drawing the blocks of first filter :
	node at (0,0)[name=input1]{$\mu$}
	node [below=1.8cm of input1](input2){$s$}
	node [output,right of=input1] (split) {}
%	node [input, name=input1] {} 
%	node [sum, right of=input1] (suma1) {\suma}
	node [block, below of=split] (cap) {\eqref{eq:Capital}}
	node [sum, right of=cap] (suma2) {\suma}
	node [block,below=0.8cm of suma2] (exo) {$\eqref{eq:Population} \quad \eqref{eq:TFP} \quad \eqref{eq:EI}$}
 	node [block, right of=suma2] (car) {\eqref{eq:Carbon}}
	node [sum, right of=car] (suma3) {\suma}
 	node [block, right of=suma3] (cli) {\eqref{eq:Climate}}
	node [above=0.8cm of suma2] (ELin) {$E_{\rm Land}$}
	node [above=0.8cm of suma3] (Fex) {$F_{\rm EX}$}
	node [output, right=0.8cm of cli] (right_out){}
	node [output, below=2.0cm of cap](left_in_dummy){}
	node [output, left=2.0cm of left_in_dummy](left_in){}
;
 %       node at (6.8,0)[block] (Q1) {\Large $Q_1$}
  %       node [block, below of=cap] (ret1) {\Large$T_1$};
    % Joining blocks. 
    % Commands \draw with options like [->] must be written individually
	\draw[->](input1) -| node {} (car.north);
%	\draw[->](input1.east) |- node{}(cap.north);
 %	\draw[->](suma1) -- node {} (cap);
	\draw[->](cap) -- node {} (suma2);
	\draw[->](suma2) -- node {} (car);
	\draw[->](car) -- node {} (suma3);
	\draw[->](suma3) -- node {} (cli);
	%\draw[->](cli) -| node[near end]{} (suma1);
	\draw[-](cli) -- node{}(right_out);
	\draw[-](right_out.south) -- ++(0,-3.2cm) -| node{} (left_in);
	\draw[->](exo.west) -| node {}(cap);
	\draw[->](exo) -| node {} (car.south);
	\draw[->](ELin) -- node {} (suma2);
	\draw[->](Fex) -- node {} (suma3);
	\draw[->](split) -- node {} (cap);
	\draw[->](input2) -- node {} ([yshift=0.25cm]cap.west);
	\draw[->](left_in.north) |- node {}([yshift=-0.25cm]cap.west);
\end{tikzpicture}
\caption{Block Diagram of DICE.}
\label{fig:BlockDiagram}
\end{center}
\hrulefill
\vspace*{4pt}
\end{figure*}
\begin{figure*}[!h]
\normalsize
\newcounter{MYtempeqnctr}
\setcounter{MYtempeqnctr}{\value{equation}}
\setcounter{equation}{0}
\begin{align}
 	T(i+1) & = \Phi_T T(i) + B_T \left( F_{2\times} \log_2 \left(\tfrac{M_{\rm AT}(i)}{M_{\rm AT,  1750}}\right) 
			+ F_{\rm EX}(i) \right) , \tag{CLI}
			\label{eq:Climate} \\	
	M(i+1) & = \Phi_M M(i) + B_M \left(\sigma(i) (1 - \mu(i)) A(i) K(i)^\gamma L(i)^{1 - \gamma} + 
			E_{\rm Land}(i)\right) , \tag{CAR}
			\label{eq:Carbon} \\
	K(i+1) & = \Phi_K K(i) 
	 		+ \Delta \left(\frac{1}{1+a_2 \, T_{\rm AT}(i)^{a_3}}\right) 
			%\left(1 - \tfrac{p_b}{1000\theta_2}(1 - \delta_{pb})^{i-1} \sigma(i) \mu(i)^{\theta_2}\right) 
			\left(1 - \theta_1(i) \mu(i)^{\theta_2}\right) 
			A(i) K(i)^\gamma L(i)^{1 - \gamma} s(i) , \tag{CAP}
		\label{eq:Capital} \\
	L(i+1) & = L(i) \left(\frac{\CMK{1+ } L_a}{\CMK{1 + } L(i)}\right)^{\ell_g},  \tag{POP} \label{eq:Population} \\
	A(i+1) & = \frac{A(i)}{1 - g_A\exp(- \delta_A  \Delta  (i-1))},  \tag{TFP} \label{eq:TFP} \\
	\sigma(i+1) & = \sigma(i) \exp\left(-g_\sigma  (1-\delta_\sigma)^{\Delta(i-1)}  \Delta \right),
		\tag{EI} \label{eq:EI}
\end{align}
\setcounter{equation}{\value{MYtempeqnctr}}
\hrulefill
\vspace*{4pt}
\end{figure*}

The dynamics of the \DICE IAM \cite{DICEManual} are given by equations \eqref{eq:Climate}--\eqref{eq:EI} and the inter-relationships shown in Figure~\ref{fig:BlockDiagram}.  \CMK{Note that the equation labels are descriptive, with 
\eqref{eq:Climate} describing the climate (or temperature) dynamics; \eqref{eq:Carbon} describing the carbon cycle
dynamics; \eqref{eq:Capital} describing capital (or economic) dynamics; \eqref{eq:Population} providing population
dynamics; \eqref{eq:TFP} giving the dynamics of total factor productivity; and \eqref{eq:EI} describing the emissions
intensity of economic activity.}

We describe each of the modeling blocks in Figure~\ref{fig:BlockDiagram} in Sections~\ref{sec:Exogenous}--\ref{sec:CarbonModel} below.  \CMK{Here, however, we note that the model is nonlinear and time-varying.}
The model assumes two control inputs: the savings rate $s$ and the mitigation or abatement rate $\mu$.  The first of these we describe more fully in Section~\ref{sec:EconomicModel}.  The latter is the rate at which mitigation of industrial carbon dioxide emissions occurs. 

The model uses a time-step of 5 years, starting in the year 2015 for DICE2016 (or 2010 for DICE2013).  Take the discrete
time index $i \in \mathbb{N}$, the sampling rate $\Delta = 5$, and the initial time
$t_0 = 2015$ (or 2010) so that
\begin{align}
	\label{eq:time_step}
	t = t_0 + \Delta \cdot (i-1)
\end{align}
and hence $t \in \{ 2015, 2020, 2025, \ldots\}$.

\subsection{DICE ``Exogenous'' States}
\label{sec:Exogenous}

The states for population $L$, total factor productivity $A$ (which is a measure of technological progress), and 
carbon intensity of economic activity $\sigma$, are 
frequently referred to as ``exogenous variables''.  This is due to the fact that they are not influenced
by the states for climate, carbon, or capital, which are frequently referred to as ``endogenous variables''
(see Figure~\ref{fig:BlockDiagram}).  

As mentioned, there are some inconsistencies in
the published literature with regards to the form of these inputs.  For the sake of
completeness and to remain close to the numerical results generated
by \cite{DICECode}, we present and use the exogenous states as defined 
in \cite{DICECode}.  Background information on the parameters and functional
form of these expressions can be found in \cite{Nordhaus_Notes2007},
\cite{Nordhaus-2008}, and \cite{DICEManual}.

\CMK{The population model \eqref{eq:Population} is referred to as the Hassell Model \cite{Hassell-1975}. 
Total factor productivity \eqref{eq:TFP} yields a logistic-type function; i.e., the total factor productivity
is monotonically increasing with a decreasing growth rate.
Carbon intensity of economic activity \eqref{eq:EI} is similar to total factor productivity in that it is a
monotonically decreasing function with a decreasing decrease rate.} 
%\begin{align}
%	L(i+1) & = L(i) \left(\frac{L_a}{L(i)}\right)^{\ell_g},  \label{eq:pop_i}\\
%	A(i+1) & = \frac{A(i)}{1 - g_A\exp(- \delta_A  \Delta  i)},   \label{eq:tfp_i} \\
%	\sigma(i+1) & = \sigma(i) \exp\left(-g_\sigma\cdot  (1-\delta_\sigma)^{\Delta(i+1)}  \Delta \right) \label{eq:ei_i},
%\end{align}
The quantities $L(1) = L_0$ and $A(1) = A_0$ are prescribed initial conditions for the global population and
total factor productivity in the base year.  

An estimate for the initial emissions intensity of economic activity
$\sigma(1) = \sigma_0$ can be calculated as the ratio of global industrial emissions to global economic
output.  The estimate of $\sigma_0$ can be further refined by estimating the mitigation rate in the base
year.  In other words, with base year emissions $e_0$, base year economic output $q_0$, and an
estimated base year mitigation rate $\mu_0$, we can estimate $\sigma_0 = \frac{e_0}{q_0 (1 - \mu_0)}$.

An estimate of the cost
of mitigation efforts is given by
\begin{equation}
	\label{eq:theta1}
	\theta_1(i) = \frac{p_b}{1000\cdot \theta_2} (1-\delta_{pb})^{i-1} \cdot \sigma(i) .
\end{equation} 
Here, $p_b$ represents the price of a backstop technology
that can remove carbon dioxide from the atmosphere.  Note that this 
equation embeds the assumption that the cost of such technology will 
decrease over time (since $\delta_{pb} \in (0,1)$) and will be proportional
to the emissions intensity of economic activity.

The remaining two exogenous signals are given by 
\begin{align}
	F_{\rm EX}(i) & = f_0 + \min \left\{ f_1 - f_0, \frac{f_1 - f_0}{t_f}(i-1) \right\} , \label{eq:FEX} \\
	E_{\rm Land}(i) & = E_{L0} \cdot (1-\delta_{EL})^{i-1}. \label{eq:eland}
\end{align}
The signals $F_{\rm EX}$ and $E_{\rm Land}$ are estimates of the effect of
greenhouse gases other than carbon dioxide and the emissions due to land
use changes, respectively.  

Numerical values for all parameters can be found in the appendix as well as in
the accompanying code \cite{Faulwasser-MPC-DICE-IAMES2018}.

\subsection{Economic Model}
\label{sec:EconomicModel}
We now turn to the economic component of the \DICE integrated assessment model.
In summary, the \DICE model assumes a single global economic ``capital''.  Capital depreciates
and is replenished by investment.  The amount available to invest is some fraction
of the net economic output which can be derived from the gross economic output.
This is a standard economic growth model (see \cite{Acemoglu09} for a comprehensive
treatment of such models).

Gross economic output is the product of three terms; the total factor productivity $A$; capital $K$; and labor $L$
approximated by the global population.  Additionally, capital and labor contribute at different levels given by 
a constant called the
capital elasticity $\gamma \in [0,1]$; that is, gross economic output is given by\footnote{\CMK{The quantity 
in \eqref{eq:output} is referred to as a \textit{Cobb-Douglas Production Function with Hicks-Neutral technological
progress} (see \cite[p.~36, p.~58]{Acemoglu09}).}}
\begin{align}
	\label{eq:output}
	Y(i) = A(i) K(i)^\gamma L(i)^{1 - \gamma}.
\end{align}
Note the expression corresponding to gross economic output in \eqref{eq:Carbon}
(see also \eqref{eq:Emission} below).
%Units: Capital, $K$, in trillions 2005 USD; Population, $L$, in millions people, normalized to billions
%people; Total Factor Productivity, $A$, hence in per billions people.

Net economic output, $Q$, is gross economic output, $Y$, reduced by two factors: 1) climate damages 
from rising atmospheric temperature, and 2) the cost of efforts towards mitigation:
\begin{align}
	Q(i) & = \left(\frac{1}{1+a_2 \, T_{\rm AT}(i)^{a_3}}\right) \left(1 - \theta_1(i) \mu(i)^{\theta_2}\right) Y(i) \nonumber \\
	& = \left(\frac{1}{1+a_2 \, T_{\rm AT}(i)^{a_3}}\right) \left(1 - \theta_1(i) \mu(i)^{\theta_2}\right) 
		 \nonumber \\
	& \qquad \cdot A(i) K(i)^\gamma L(i)^{1 - \gamma},
	\label{eq:Net_Economic_Output}
\end{align}
where $\theta_1$ is as defined in \eqref{eq:theta1}.
Observe the component of net economic output in \eqref{eq:Capital}.
%Units: Net Economic Output, $Q$, in trillions 2005 USD.  $\mu$ dimensionless and, hence, so is
%$\theta_1$.   $0.00267$ must be in per $^\circ$C.

Net economic output can then be split between consumption and investment
\begin{align}
	\label{eq:OutputSplit}
	Q(i) = C(i) + I(i)
\end{align}
and the savings rate is defined as
\begin{align}
	\label{eq:SavingsRate}
	s(i) = \frac{I(i)}{Q(i)}.
\end{align}
%With investment, $I$, in trillions 2005 USD, $s$ is dimensionless.

The economic dynamics \eqref{eq:Capital} are a capital accumulation model where capital depreciates 
according to
\begin{equation}
	\label{eq:PhiK}
	\Phi_K \doteq (1 - \delta_K)^\Delta
\end{equation}
and is replenished by investment $I$ in the form of the product of the savings rate and 
net economic output; i.e.,
\begin{align}
	K(i+1) & = \Phi_K K(i) + \Delta \cdot I(i) \nonumber \\
	& = \Phi_K K(i) +  \Delta \cdot Q(i) s(i) 	\label{eq:Capital_Intermediate}
\end{align}
and substitution of \eqref{eq:Net_Economic_Output} into \eqref{eq:Capital_Intermediate}
yields \eqref{eq:Capital}.  The savings rate $s$ is the second of the two control inputs.

\subsection{The Damages Function}
One of the two most contentious elements\footnote{The other being the discount rate discussed below.} 
in climate economics is the specification of the
damages function (see \cite{Bonen14}).  This stems from the
inherent difficulty of modeling in an application where experimentation is 
simply not possible and the fact that rising temperatures will have different
local effects.  Hence, different researchers have proposed many different
damage functions and levels \cite{Bonen14}.  The specific form of the damages function in
DICE, as shown in \eqref{eq:Capital}, is
\begin{equation}
	\label{eq:Damages}
	\frac{1}{1+a_2 T_{\rm AT}(i)^{a_3}}
\end{equation}
where, with $a_3 = 2$, the parameter $a_2$ is calibrated to yield a loss of 2\% at 3 $^\circ$C
(see \cite{Nordhaus-NBER2017} for the calibration of this and other parameters).

While a full discussion of the appropriateness of \eqref{eq:Damages} is beyond the 
scope of this article, it is worthwhile noting that it has been vigorously
argued that the above calibration of 2\% loss at 3 $^\circ$C is unreasonably low if it is to be
consistent with currently available climate science \cite{Stern-JEconLit-2013}.  
\CMK{Recent efforts to empirically estimate climate damages can be found in
\cite{Hsiang-etal-2017}.}

\subsection{Climate Model}
\label{sec:ClimateModel}
The climate or temperature dynamics used in \DICE are derived from a two-layer energy balance model \cite{Geoffroy2013,Gregory2000,Schneider1981Geophysic}.
In particular, a simple \CMK{explicit} Euler discretization is applied to an established continuous-time energy balance model
to obtain \eqref{eq:Climate}.  However, the implementation in \cite{DICECode} \CMK{is not strictly causal in that
the atmospheric temperature at the next time step depends on the radiative forcing at the next time step.
This was previously observed in \cite{Cai-Judd-Lontzek-2012} and \cite{Calel-Stainforth-Dietz-2015}.
We explicitly provide the derivation of the model here for future reference.  Furthermore, we note that
using the causal model below, as opposed to the version in \cite{DICECode}, has a negligible quantitative impact on the numerical results obtained using the model.}

The two layers in the energy balance model are the combined atmosphere, land surface, and upper ocean (simply 
referred to as the atmospheric layer in what follows) and the lower ocean.  We denote these two states by
$T_{\rm AT}$ and $T_{\rm LO}$, respectively, and the zero reference is taken as the temperature in the year 1750.  
With $F(t)$ denoting the radiative forcing at the top of atmosphere
due to the enhanced greenhouse effect,
the (continuous-time) dynamics for these states are given by
\begin{subequations} \label{eq:ct_climate}
\begin{align}
	C_{\rm AT} \frac{d \, T_{\rm AT}(t)}{dt} & = F(t) - \lambda T_{\rm AT}(t) \nonumber \\
		& \qquad - \gamma \left(T_{\rm AT}(t) - T_{\rm LO}(t) \right) , \label{eq:atm_ct} \\
	C_{\rm LO} \frac{d \, T_{\rm LO}(t)}{dt} & =\gamma \left(T_{\rm AT}(t) - T_{\rm LO}(t) \right).
\end{align}
\end{subequations}
Here, $C_{\rm AT}$ and $C_{\rm LO}$ are the heat capacities of the atmospheric and lower ocean layers and $\gamma$
is a heat exchange coefficient.  The quantity $\lambda$ is called the \emph{Equilibrium Climate Sensitivity (ECS)}.  We discuss the ECS in Remark~\ref{rem:lambda} below.

Taking an \CMK{explicit} Euler discretization with time-step $\Delta$ yields
\begin{subequations}
\begin{align}
	T_{\rm AT}(i+1) & = T_{\rm AT}(i) + \frac{\Delta}{C_{\rm AT}} \left( F(i) - \lambda T_{\rm AT}(i) \right. \nonumber \\ 
		& \qquad \left. - \gamma \left(T_{\rm AT}(i) - T_{\rm LO}(i) \right) \right) \label{eq:atm_dt}\\
	T_{\rm LO}(i+1) & = T_{\rm LO}(i) \nonumber \\
		& \qquad + \frac{\Delta}{C_{\rm LO}} \left( \gamma \left( T_{\rm AT}(i) - T_{\rm LO}(i) \right) \right).
\end{align}
\end{subequations}
With $T \doteq [ T_{\rm AT} \ T_{\rm LO}]^\top \in \mathbb{R}^2$, the above\footnote{\CMK{
The implementation in \cite{DICECode} replaces $F(i)$ with $F(i+1)$ in \eqref{eq:atm_dt}.  
This could be interpreted as an implicit Euler discretization.  However, 
an implicit Euler discretization of \eqref{eq:ct_climate} would lead to very different 
expressions for the constants in \eqref{eq:Climate_params}.}} can be rewritten as 
\begin{align}
	T(i+1) = \Phi_T T(i) + B_T F(i) 
\end{align}
where
\begin{subequations} \label{eq:Climate_params}
\begin{equation}
	\Phi_T \doteq \left[ \begin{array}{cc} \phi_{11} & \phi_{12} \\ \phi_{21} & \phi_{22} \end{array} \right],
	\qquad B_T \doteq \left[ \begin{array}{c} \xi_1 \\ 0 \end{array} \right] 
\end{equation}
and
\begin{align}
	\phi_{11} & = 1 - \frac{\Delta}{C_{\rm AT}} ( \lambda + \gamma ) \\
	\phi_{12} & = \frac{\Delta \gamma}{C_{\rm AT}} \\
	\phi_{21} & = \frac{\Delta \gamma}{C_{\rm LO}} \\
	\phi_{22} & = 1 - \frac{\Delta \gamma}{C_{\rm LO}} \\
	\xi_1 & = \frac{\Delta}{C_{\rm AT}} \ . \label{eq:xi1}
\end{align}
\end{subequations}

Atmospheric temperature rise is driven by radiative forcing, or the greenhouse effect, at the top of atmosphere and, as shown in \eqref{eq:Climate} has a nonlinear (logarithmic) dependence on the mass of CO$_2$ in the atmosphere, $M_{\rm AT}$.  Greenhouse gases other than CO$_2$ (e.g., methane, nitrous oxide, and chloroflourocarbons) contribute to the radiative forcing effect, and these are accounted for in the \DICE model by the exogenously defined signal, $F_{\rm EX}$ in \eqref{eq:FEX}.  
\begin{remark}[Equilibrium Climate Sensitivity]
	\label{rem:lambda}~\\
	The parameter $\lambda$ in \eqref{eq:atm_ct} has a specific physical interpretation in terms of the radiative 
	forcing and an experiment involving the doubling of atmospheric carbon.  
	Let $F_{2\times}  > 0$ denote the forcing associated with equilibrium carbon doubling.
	Ignoring the contribution of the exogenous forcing $F_{\rm EX}$, the radiative
	forcing is given by
	\begin{align}
		\label{eq:forc}
		F(t) = F_{2\times} \log_2 \left( \frac{M_{\rm AT} (t)}{M_{\rm AT, \, 1750}} \right)
	\end{align}
	where $M_{\rm AT, \, 1750}$ is the atmospheric mass of carbon in the year 1750.  Doubling the 
	value of atmospheric carbon from pre-industrial equilibrium yields
	radiative forcing of
	\begin{align}
		F_{2\times} \log_2 \left( \frac{2 M_{\rm AT, \, 1750}}{M_{\rm AT, \, 1750}} \right) = F_{2\times} .
	\end{align}
	%Note that in \eqref{eq:Climate}, $F_{2\times}$ is denoted by $\eta$.
	
	The Equilibrium Climate Sensitivity (ECS) is defined as the steady-state atmospheric temperature arising from a doubling of atmospheric
	carbon.  Hence, for thermal equilibrium corresponding to a doubling of atmospheric carbon, we can combine \eqref{eq:atm_ct}
	and \eqref{eq:forc} to see that
	\begin{align}
		F_{2\times} \log_2 \left( \frac{2 M_{\rm AT, \, 1750}}{M_{\rm AT, \, 1750}} \right) - \lambda \, {\rm ECS} = 0
	\end{align}
	or $\lambda = F_{2 \times} / {\rm ECS}$.  In words, $\lambda$ is the ratio between the radiative forcing associated with a doubling
	of atmospheric carbon and the equilibrium atmospheric temperature arising from such a doubling.
\end{remark}

\subsection{Carbon Model}
\label{sec:CarbonModel}
Similar to the temperature dynamics, \eqref{eq:Carbon} is a three-reservoir model of the global carbon cycle, with states describing the average mass of carbon in the atmosphere, $M_{\rm AT}$, the upper ocean, $M_{\rm UP}$, and the deep or lower ocean, $M_{\rm LO}$.  We denote the carbon states by $M \doteq [M_{\rm AT} \ M_{\rm UP} \ M_{\rm LO}]^\top \in \mathbb{R}^3$ and the coefficients $\zeta_{ii} \in [0,1]$ give the diffusion between reservoirs.  We define
\begin{equation}
	\label{eq:Carbon_params}
	\Phi_M \doteq \left[ \begin{array}{ccc} \zeta_{11} & \zeta_{12} & 0 \\ \zeta_{21} & \zeta_{22} & \zeta_{23} \\
			0 & \zeta_{32} & \zeta_{33} \end{array} \right], \qquad
	B_M \doteq \left[ \begin{array}{c} \xi_2 \\ 0 \\ 0 \end{array} \right] .
\end{equation}

The mass of atmospheric carbon is driven by CO$_2$ emissions due to economic activity\footnote{Note that the parameter $\xi_2$ is simply for converting CO$_2$ to carbon (see Appendix B).}.  This occurs via a nonlinear, time-varying function as shown in \eqref{eq:Carbon}, that corresponds to modeled predictions of emissions and the emissions intensity of economic activity.  The additional term, $E_{\rm Land}$, captures emissions due
to land use changes as given by \eqref{eq:eland} above.  Hence, the total emissions are described by
\begin{multline} \label{eq:Emission}
E(i) = \\ \sigma(i) (1 - \mu(i)) A(i) K(i)^\gamma L(i)^{1 - \gamma} + 
			E_{\rm Land}(i).
\end{multline}

Note that this model is conceptually similar to the three reservoir model of the 
Global Carbon Budget project \cite{GlobalCarbonBudget2018}.  However,
the three reservoirs used by the Global Carbon Budget correspond to 
atmospheric, ocean, and land reservoirs.

\subsection{Welfare Maximization}
\label{sec:WM}
While the nine state, two decision variable \DICE model \eqref{eq:Climate}--\eqref{eq:EI}
can be used to predict outcomes based on externally (e.g., expert) predicted
mitigation and savings rates, the inputs, and predicted outcomes, are more usually
the result of solving an Optimal Control Problem (OCP).  In particular, the \DICE dynamics
act as constraints in a social welfare maximization problem.

The social welfare $W$ is defined as the discounted sum of (time-varying) utility $U$ which 
depends on consumption.  Consumption \CMK{is derived} from \eqref{eq:OutputSplit}
and \eqref{eq:SavingsRate} as 
\begin{equation} \label{eq:Consumption}
C(i) = Q(i)(1-s(i))
\end{equation}
  which
can be written explicitly in terms of states and inputs using \eqref{eq:Net_Economic_Output}.

The utility is taken as:
\begin{align}
	\label{eq:Utility}
	U(C(i),L(i)) = L(i) \left( \frac{\left(\frac{C(i)}{L(i)}\right)^{1-\alpha} - 1}{1 - \alpha} \right),
\end{align}
where $\alpha \geq 0$ is called the elasticity of marginal utility of consumption.  Note that, in the limit as 
$\alpha \rightarrow 1$, the utility is a logarithmic function of per capita consumption \CMK{and for $\alpha \in (0,1)$,
\eqref{eq:Utility} behaves qualitatively like a logarithm.  For $\alpha > 1$, since the population $L(i)$ is bounded
(as the solution of \eqref{eq:Population}), the utility is also bounded.  Indeed, denoting the upper bound on population
by $\widehat{L}$ (i.e., $L(i) \leq \widehat{L}$ for all $i$), we see that
\begin{align*}
	\lim_{(C/L) \rightarrow \infty} L \left( \frac{\left(\frac{C}{L}\right)^{1-\alpha} - 1}{1 - \alpha} \right)
	\leq - \frac{\widehat{L}}{1-\alpha} .
\end{align*}
}

The optimal control problem of interest, maximizing the social welfare, is then given by
\begin{align}
	W^\star \doteq  \max_{{\bf s}, {\bf \mu}}& \sum_{i=0}^{\infty} \frac{U(C(i),L(i))}{(1+\rho)^{\Delta i}}
	\nonumber\\
	 {\rm subject \ to} &  \tag{\rm OCP} \label{eq:OCP} \\
	& \eqref{eq:Climate}-\eqref{eq:EI}, \eqref{eq:FEX}-\eqref{eq:eland} \nonumber \\
	& %\hspace*{1.6cm} 
	\mu(i), s(i) \in [0,1], \forall i \in \mathbb{N}, \nonumber
\end{align}
where $\rho > 0$ is a prescribed discount rate.

\begin{remark}[Dis. rates and soc. time preference]
It should be mentioned that the numerical value chosen for $\rho$ can have a significant impact
on the results and is a subject of significant discussion.
  On the one hand, when analyzing capital 
investment decisions, discount rates of 7-9\% are common \cite{OMB-Circular}.  On the other hand, given the extremely
long time scales involved in the climate system and hence the long-term impacts of current emissions,
the discount rate can also be viewed in the context of intergenerational fairness (or social time preference). 
Here, arguments have been made for an effective 0\% discount rate \cite{Stern-JEconLit-2013}, though rates of 1-3\% 
are more common.   In the results that follow, when not otherwise specified, we use the default value of 1.5\% as in \cite{DICECode}.
\end{remark}

\subsection{The Social Cost of Carbon Dioxide (SC-CO$_2$)}
\label{sec:SCC}
\begin{figure}[t]
	\centering
	\includegraphics[clip, trim= 0.95cm 0.0cm 0.95cm 0.0cm, width=0.45\textwidth]{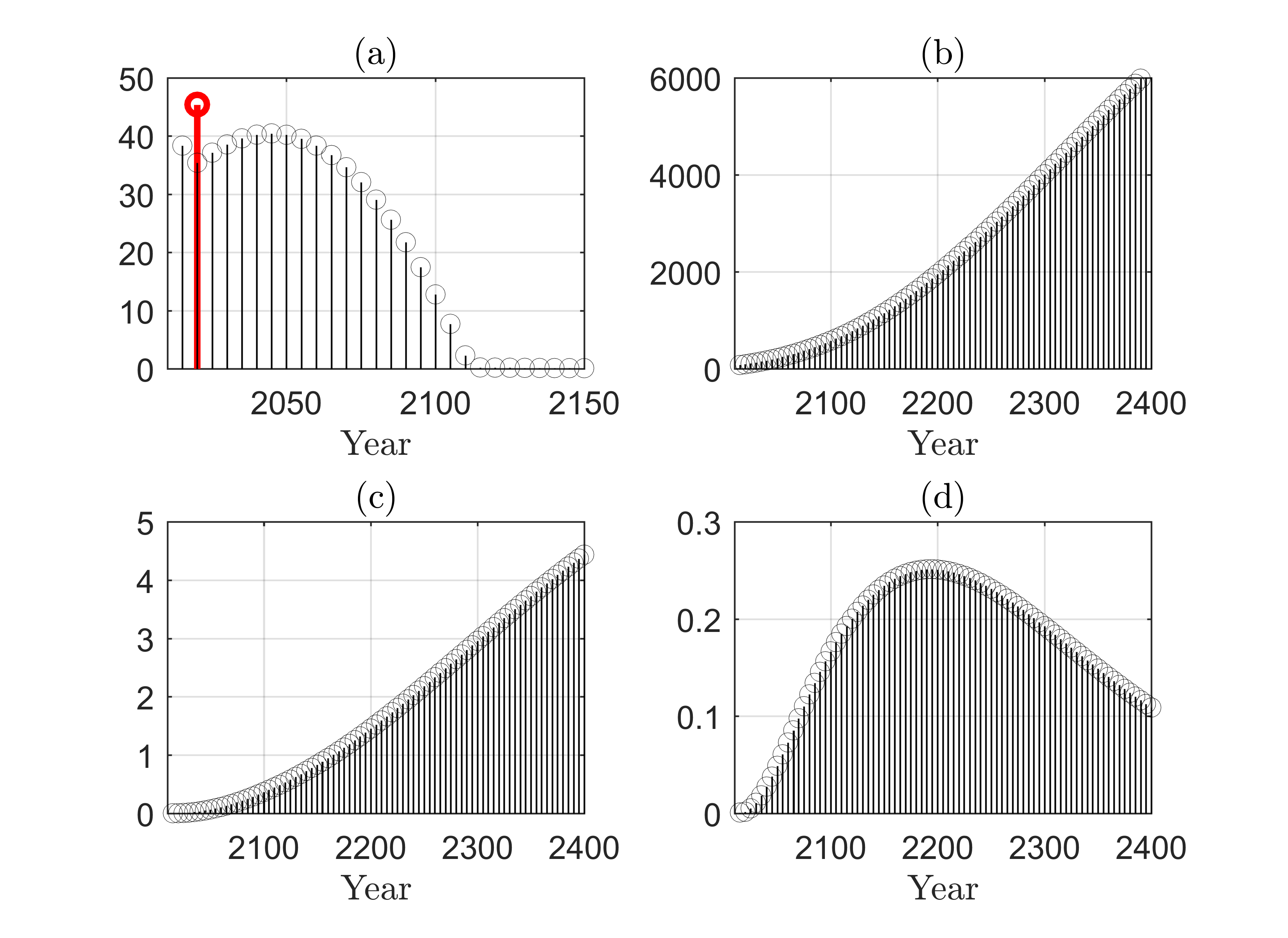}
	\caption{\CMK{The Social Cost of Carbon Dioxide (\SCC) computed from an emissions pulse
	experiment.  (a) Baseline emissions and a pulse in the year 2020 (vertical axis in gigatonnes CO$_2$).  
	(b) Consumption pathways resulting from the baseline and pulse emissions pathways are virtually
	indistinguishable (vertical axis in millions of 2010USD).  (c) The difference between the baseline 
	consumption pathway and the emissions pulse pathway (vertical axis in millions of 2010USD).
	(d) The 5\% discounted difference between the two consumption pathways.  Summing the values in
	this plot, and normalizing by the size of the emissions pulse, yields the \SCC for 2020.}}
	\label{fig:SCCcalc}
\end{figure}

%\piccaptioninside
%\piccaption{Social cost of carbon.   
%\label{fig:SCCcalc}}
%\parpic[fr]{\fbox{\includegraphics[width=7.0cm]{SCCmanualapproach.png}
%			%\caption{}
%			}}
The \TF{Social Cost of Carbon Dioxide} (\SCC) in a particular year is defined as 
\begin{quote}
``the decrease in aggregate consumption in that year that would change the current \ldots value of social welfare by the same amount as a one unit increase in carbon emissions in that year.'' \cite{Newbold-Rapid-Assessment-2013}
\end{quote}  
 \CMK{This can be computed as shown in Figure~\ref{fig:SCCcalc} where baseline emissions and consumption (Figure~\ref{fig:SCCcalc} (a) and (b)) are defined, e.g., by solving the optimal control problem \eqref{eq:OCP}.  A pulse of CO$_2$ emissions is then injected at a particular year (10 GtCO$_2$ in 2020 in Figure~\ref{fig:SCCcalc}(a)) and the aggregate reduction in consumption (Figure~\ref{fig:SCCcalc}(c)) over succeeding years, appropriately discounted (Figure~\ref{fig:SCCcalc}(d)), is the \SCC for that year. 
Note the different time scales between Figure~\ref{fig:SCCcalc}(a) and \ref{fig:SCCcalc}(b)-(d), which 
emphasizes that although industrial emissions in this scenario go to zero shortly after the year 2100,
the effects of these emissions, even discounted, persist far into the future.} 

These pulse experiments are suggestive of a sensitivity analysis and, in fact, the computation of
the \SCC is given by the ratio of two Lagrange multipliers.  Specifically, the Lagrange 
multipliers of interest are the incremental change in welfare with respect to the incremental change in 
emissions, $\frac{\partial W^\star}{\partial E(i)}$, and the incremental change in welfare with respect to the incremental change in consumption, $\frac{\partial W^\star}{\partial C(i)}$. 
The \SCC is then given by
\mathchardef\mhyphen="2D
\begin{align}
	{\rm SC \mhyphen CO_2}(i) & = - 1000 \cdot \frac{\partial W^\star / \partial E(i)}{\partial W^\star / \partial C(i)} \nonumber \\
	& = - 1000 \cdot \frac{\partial C(i)}{\partial E(i)} .
	\label{eq:SCC}
\end{align}
Note that the factor of 1000 scales the \SCC to 2010 US dollars per tonne of CO$_2$, whereas
consumption is in trillions of 2010 US dollars and emissions are in gigatonnes of CO$_2$.

As mentioned above, the discount rate $\rho$ can have a significant impact on the optimal solution and, hence, the monetary value of emissions given by the \SCC.  Table \ref{tab:SCC} lists the \SCC for different years and the different discount rates $\rho\in\{0.005,\,0.015,\, 0,03\}$ and considering a finite horizon $N=100$. For example, the estimates \SCC for the year 2020 range from US\$12.55 ($\rho = 0.03$) to
US\$89.31 ($\rho = 0.005$).
\begin{table}[t]
\caption{\SCC computed for DICE2016 via \cite{DICECode} for selected years and different values of the discount rate $\rho$. \label{tab:SCC}}
\begin{center}
\begin{tabular}{|c||r|r|r|} \hline
	Year	& $\rho = 0.005$		& $\rho = 0.015$		& $\rho = 0.03$ \\ \hline\hline
2015	& US\$73.95	& US\$27.14	& US\$10.84 \\ \hline
2020	& US\$89.31	& US\$32.28	& US\$12.54 \\ \hline
2030	& US\$124.20	& US\$44.54	& US\$16.98 \\ \hline
\end{tabular}
\end{center}
\end{table}

\subsection{\DICE, \PAGE, and \FUND}
\label{sec:PAGEFUND}
While many integrated assessment models have been proposed,
the three most commonly used and cited models are DICE, \PAGE,
and \FUND.  In particular, the U.S. Interagency Working Group
made use of these three models in deriving its estimates of the 
\SCC \cite{InteragencyWorkingGroup}.  The \PAGE model was
used extensively in the Stern Review \cite{SternReview}.

PAGE and \FUND are fundamentally different models than 
DICE.  In the economics lexicon, \PAGE and \FUND are ``partial equilibrium
models'' while \DICE is a ``general equilibrium model''.
Specifically, economic growth is an input in the former type of model, but
a state (given by the evolution of $K$) in the latter.  As a consequence,
in solving the welfare maximization problem \eqref{eq:OCP},
DICE generates optimal emissions and consumption pathways.
By contrast, such pathways must be provided as inputs to 
PAGE and \FUND.

The PAGE model divides the world into eight regions and considers four different
damages components given by sea level rise, economic damages, non-economic
damages, and discontinuities.  This is in contrast to DICE which considers a single global
region and a single damages component.  Additionally, PAGE looks to incorporate uncertainty
by repeatedly drawing several parameters from probability distributions.  The model
is instantiated as an Excel spreadsheet and makes use of the proprietary 
{@}RISK software add-in \cite{ExcelRISK} to perform the required Monte Carlo
calculations.

It is important to note that PAGE takes not just economic growth (or projected Gross Domestic Product)
as an input, but also climate policies (such as the mitigation rate $\mu$) as inputs.  In other words,
there is no optimization problem associated with the model.

FUND is similar in concept to PAGE as a partial equilibrium model, but differs in its specifics.
FUND considers sixteen geographic regions and eight damages components.  Furthermore, some of
the damages components are dependent on both the temperature increase and the \textit{rate}
of temperature rise or CO$_2$ concentrations, while damages in DICE and PAGE are dependent
solely on the temperature increase.  FUND is coded in C\# and is available at \cite{FundModel}.

Given the fact that neither PAGE nor FUND involve an optimal control problem, they do not
compute the \SCC as per \eqref{eq:SCC}, but rather do so via the pulse experiment 
as indicated in Figure~\ref{fig:SCCcalc}.

Finally, we note that a regional variant of DICE---called RICE (Regional Integrated model of Climate
and Economy)---was considered by Nordhaus in conjunction with the 2010 variant of DICE
\cite{Nordhaus-PNAS2010}.  The RICE model used the same geophysical structure as previously
described for DICE, but considered twelve global regions by calibrating twelve essentially independent
copies of the economic model \eqref{eq:Capital}.

\section{Receding Horizon Solution to DICE}
\label{sec:RHC}

As defined, \eqref{eq:OCP} is a \CMK{non-convex} infinite-horizon \CMK{optimal control} problem and is thus difficult to solve analytically and numerically.\footnote{\CMK{Note that \cite{DICECode} solves a slightly different problem than \eqref{eq:OCP}.  Specifically, \cite{DICECode} solves over a fixed horizon of 60 or 100 (corresponding to 300 or 500 years), and fixes the savings rate over the last ten time steps to a value close to the turnpike value.  This latter element precludes the capital stock from being depleted at the end of the fixed horizon.  Conceptually, fixing the horizon {\it a priori} rules out discount rates below a certain threshold since numerically significant behavior occurs on long time scales but is not rendered insignificant by the discounting.}} However, from a systems-and-control perspective it is intuitive to approximate the solution to the  infinite-horizon problem \eqref{eq:OCP} by means of a receding-horizon---or model predictive control---approach. Recently the analysis of asymptotic properties of model predictive control with generic objective functionals (that do not explicitly encode a control task)  has received significant attention under the label \textit{economic MPC}, cf. \cite{Rawlings17,kit:faulwasser18c}. Indeed, \CMK{for a very general class of problems} in the undiscounted time-invariant setting, it can be shown that the receding-horizon approach yields a quantifiably accurate approximation of the infinite-horizon solution that improves as the horizon increases\footnote{\CMK{Specifically, the analysis of economic MPC schemes leverages so-called \emph{turnpike properties} of OCPs. Turnpike properties are similarity properties of parametric OCPs, whereby for varying horizon lengths and varying initial conditions, the time that the solution spends close to a specific attractor---i.e., close to the turnpike---grows with horizon length. Early observations of this phenomenon can be traced back to John von Neumann \cite{vonNeumann38}, while the term ``turnpike'' was coined in 1958 in \cite{Dorfman58}. The concept has received widespread attention in economics \cite{Mckenzie76,Carlson91} and, more recently, in systems and control \cite{Trelat15a,Gruene16a,epfl:faulwasser15h}.}} \cite{Grun16-DMV}.
\CMK{Extending the approximation results of \cite{Grun16-DMV} to include time-varying systems and discounted optimal control 
problems is the subject of ongoing work and some specific indications are provided at the end of this section and in Section~\ref{sec:Summary}.}

It is worth noting that, despite the familiarity of economists with optimal control methods (e.g., \cite{SeierstadSydsaeter}) and despite its significant impact on systems and control, the receding-horizon approach is still largely unknown in the economics community \cite{GrueneNMPCforecon}. However, from a control point of view the welfare maximization described in \eqref{eq:OCP} immediately suggests a receding-horizon approach for at least two reasons: One, it is conjectured that,
as in the case of undiscounted optimal control problems, receding 
horizon control likely provides an approximate solution to the infinite
horizon optimal control problem.  
Two, a receding horizon implementation provides a natural framework
for considering robustness issues by, in particular, separating the `plant'
and the model of the plant used for control purposes \cite{HafeezIFAC2017}.\footnote{Interestingly, in 2015 the EU called for revisiting emission reduction targets every five years \cite{Steinhauser15}, which can also be understood as a feedback mechanism.}  This idea is intuitive from a control point view, yet is not the standard for climate-economy assessment.
To the best of the authors' knowledge the earliest application of a receding-horizon framework to the \DICE OCP was presented in \cite{ChuDuncan-CDC12}, which looked at the RICE model, while \cite{WellerCDC2015} considered the DICE model and explicitly accounted for uncertainty in emissions and temperature measurements
in relation to the \SCC.

Subsequently, we aim at solving \eqref{eq:OCP} in a receding-horizon fashion to the end of computing the \SCC. As mentioned before, the \SCC definition  \eqref{eq:SCC} can be read as a quotient of Lagrange multipliers (or adjoint states). Hence, following our development in \cite{Faulwasser-MPC-DICE-IAMES2018}, we reformulate the \DICE dynamics such that the consumption $C$ and the $E$ formally can be regarded as state variables. In turn this implies that the required Lagrange \CMK{multipliers} / adjoints states are readily available upon solving \eqref{eq:OCP} \CMK{using state-of-the-art optimization codes}.

We begin by
defining the augmented state vector
\begin{align*}
 \tilde x\phantom{_{aux}} &= \begin{bmatrix}
i & T&  M&  K& \sigma&
 L & A_{\rm TFP} & E_{\rm Land} & F_{\rm EX}
  \end{bmatrix}^\top,  \\
 x_{aux}& = \begin{bmatrix} E(i) & C(i) & \mu(i) & s(i)& W(i) \end{bmatrix}^\top.
\end{align*}
Note that $\tilde x(i) \in \mbb{R}^{12}$ collects the time index $i$, the state variables of \eqref{eq:Climate}--\eqref{eq:EI}, \CMK{and the sequences \eqref{eq:FEX} and \eqref{eq:eland}}. The vector $x_{aux}(i) \in \mbb{R}^{5}$
collects the emissions \eqref{eq:Emission}, consumption \eqref{eq:Consumption}, inputs $\mu(i)$ and $s(i)$ at time $i$, and the extra state 
\[
W(i) = \sum_{j=1}^{i} \frac{U(C(j),L(j))}{(1+\rho)^{\Delta(j-1)}}
\] 
which is used to define the objective (social welfare).
Moreover, using
\[
x(i) \doteq \begin{bmatrix} \tilde x(i)^\top & x_{aux}(i)^\top\end{bmatrix}^\top
\]
and the shifted input variables
\[
w(i) \doteq \begin{bmatrix} \mu(i+1) & s(i+1)\end{bmatrix}^\top,
\]
we can rewrite the dynamics underlying \eqref{eq:OCP} as follows:
\begin{equation} \label{eq:AugmentedDyn}
x(i+1) = f(x(i), w(i)), \quad x(1) = v.
\end{equation}
The first component of the righthand-side function $f:\mbb{R}^{17}\times\mbb{R}^2 \to \mbb{R}^{17}, f = [f_1, \dots, f_{17}]^\top$ is given by
\[
f_1(x, w) \doteq
x_1+1,
\]
and the components $k = 2,\dots, 12$ are given by \eqref{eq:Climate}--\eqref{eq:EI}, \CMK{\eqref{eq:FEX},
and \eqref{eq:eland}}. 
For $k= 13$ we obtain from \eqref{eq:Emission} 
\begin{align*}
	\lefteqn{E(i+1) = f_{13}(x(i), w(i))} \quad & \nonumber \\
		& =  \Delta \left( \rule{0pt}{15pt} \sigma(i+1) (1 - \mu(i+1)) Y(i+1)\right. %\nonumber \\%A(i+1) K(i+1)^\gamma \left(\tfrac{L(i+1)}{1000}\right)^{1 - \gamma}  
		%& \phantom{==} \qquad 
		\left. + E_{\rm Land}(i+1) \rule{0pt}{15pt} \right) \nonumber\\
		&=  \Delta \left( \rule{0pt}{15pt} f_8(x(i), w(i))\cdot (1 - w_1(i)) \cdot f_{10}(x(i),w(i)) 
			\right. \nonumber \\
		& \phantom{==} \qquad \cdot f_7(x(i), w(i))^\gamma  \cdot
		 	 \left(\tfrac{f_9(x(i), w(i))}{1000}\right)^{1 - \gamma} \nonumber \\
		 & \phantom{==} \qquad  \left. + \, E_{L0} \cdot (1 - \delta_{EL})^i  \rule{0pt}{15pt}\right). 
%x_{10}(i) x_7(i)^\gamma \left(\tfrac{x_{10}(i)}{1000}\right)^{1 - \gamma}  
%		+ 3.3\cdot0.8^i \\
\end{align*}
In other words, we can rewrite the emissions explicitly as a state using 
\eqref{eq:Capital}--\eqref{eq:TFP} to expand $f_7, f_8, f_9$, and $f_{10}$.
Immediately from the above, we obtain the initial emissions
\begin{align*}
E(1) &=  \Delta \left( \rule{0pt}{15pt}x_8(1) (1 - x_{15}(1)) x_{10}(1) x_8(1)^\gamma \left(\tfrac{x_9(1)}{1000}\right)^{1 - \gamma}  \right. \nonumber \\
& \phantom{==} \qquad \left. \rule{0pt}{15pt}
		+  E_{L0} \right).
\end{align*}
Similarly, we may rewrite the consumption state equation as
\begin{align*}
	\lefteqn{C(i+1) = f_{14}(x(i),w(i))} \quad & \nonumber \\
	&= \Delta \left( \frac{1 - \theta_1(i+1) \mu(i+1)^{\theta_2}}{1 + a_2 \, T_{\rm AT}(i+1)^{a_3}} \right) %\nonumber \\
%&\phantom{==}	\qquad \rule{0pt}{15pt}	
		 Y(i+1) (1-s(i+1)) \nonumber\\
%A(i+1) K(i+1)^\gamma \left(\tfrac{L(i+1)}{1000}\right)^{1 - \gamma} 
	&= \Delta \left( \frac{1 - \theta_1(i+1) w_1(i)^{\theta_2}}{1 + a_2 \, f_2(x(i),w(i))^{a_3}} \right)  \cdot 
		f_{10}(x(i),w(i)) \nonumber \\
	&\phantom{==} \qquad   \cdot f_7(x(i), w(i))^\gamma \cdot 
		\left(\dfrac{f_9(x(i), w(i))}{1000}\right)^{1 - \gamma} \nonumber \\
	& \phantom{==} \qquad \quad \cdot (1-w_2(i))
\end{align*}
with initial condition given by
\begin{align*}
C(1) & = \Delta \left( \frac{1 - \theta_1(1) x_{15}(1)^{\theta_2}}{1+ a_2 \, x_2(1)^{a_3}}\right) 
	\cdot x_{10}(1) \cdot x_8(1)^\gamma \nonumber \\
	&  \phantom{==} \qquad \cdot
			 \left(\tfrac{x_8(1)}{1000}\right)^{1 - \gamma} \cdot (1-x_{16}(1)) .
\end{align*}
The final three states are given by
\begin{align*}
x_{15}(i+1) &= w_1(i),  \quad x_{15}(1) = v_{15}\\
x_{16}(i+1) &= w_2(i), \quad x_{16}(1) = v_{16} \\
x_{17}(i+1)&= x_{17}(i) + \frac{U(x_{12}(i),x_9(i))}{(1+\rho)^{\Delta(i-1)}}, \quad x_{17}(1) = 0.
\end{align*}
Observe that the initial condition $x_{14}(1) = C(1)$ depends on the (unshifted) inputs at time $i=1$; i.e. it depends on $\mu(1) = x_{15}(1)$ and $s(1)=x_{16}(1)$.
Likewise the initial condition $x_{13}(1) = E(1)$ depends on $\mu(1) = x_{15}(1) $. 

To handle this dependence in the optimization, we introduce the auxiliary decision variable $v \in \mbb{R}^{17}$ and  the additional constraint $x(1) = v$.
Now, we can summarize the equivalent (finite-horizon) reformulation of  \eqref{eq:OCP}  based on the augmented dynamics \eqref{eq:AugmentedDyn} as follows
\begin{subequations} \label{eq:OCP1}
\begin{align}
\max_{{\bf w}, \,v} & \ \  x_{17}(N+1) \\
\textrm{subject to } \nonumber\\
		x(j+1) &= f(x(j), w(j)),  \quad j = 1, \ldots, N  \\
		x(1) &= v\\
		  v_k &= x_k(1), \, k \in \{1,\dots, 17\}\setminus\{15, 16\}\\
  		   v_{k} &\in [0,\,1], \quad k= 15, 16   \\
		   w(j) &\in [0,\,1]\times[0,1],\quad j = 1, \ldots, N.		  		  
\end{align}
\end{subequations}

\begin{figure*}
    \centering
    \begin{subfigure}[t]{0.48\textwidth}
     %   \centering
        \includegraphics[clip, trim= 0.95cm 0.0cm 0.95cm 0.0cm, width=.9375\textwidth]{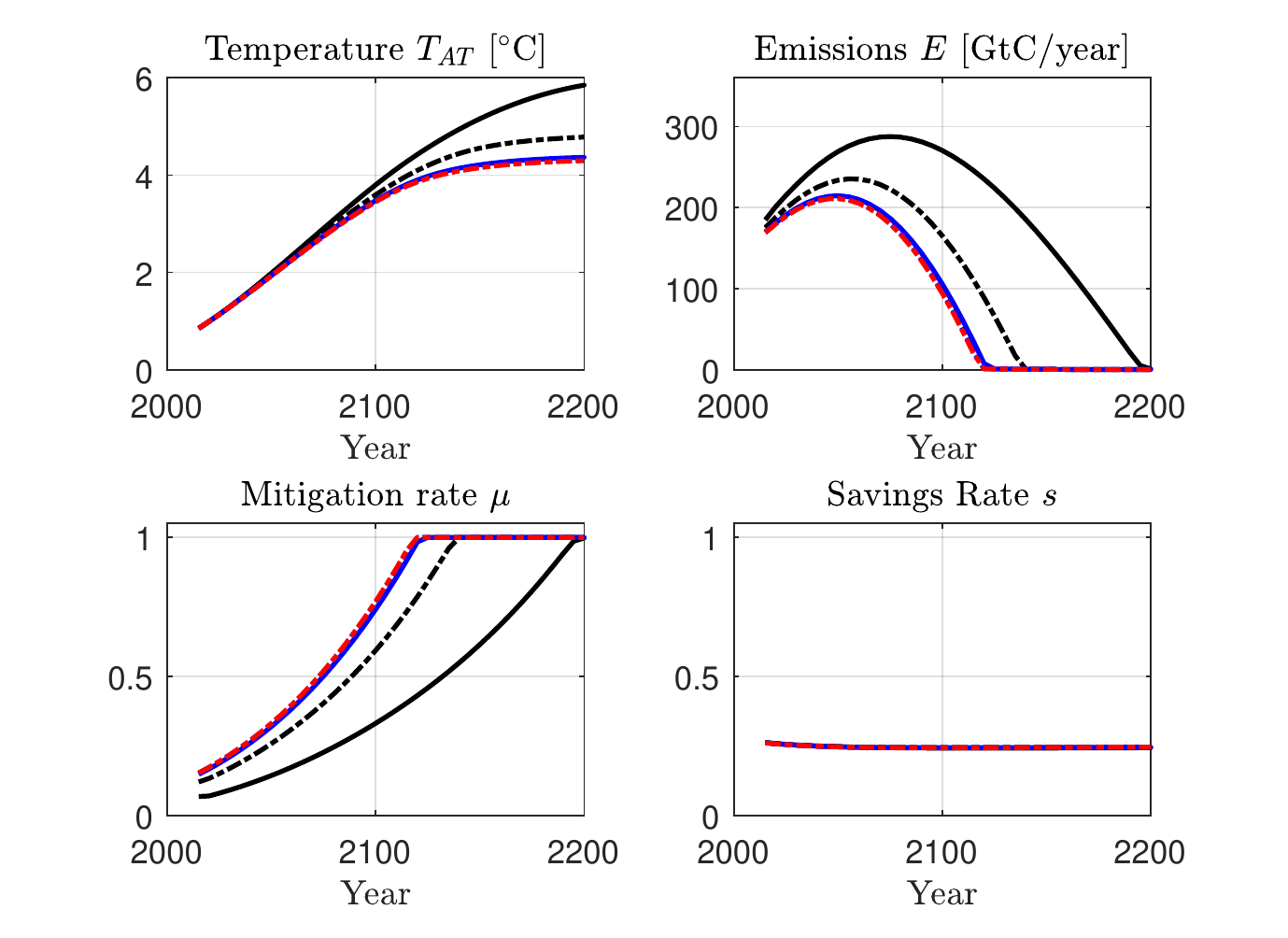}  
        \caption{%Results for \MPCDICE 2013 for varying prediction horizons using Algorithm \ref{algo:MPC-DICE} with $N \in \{10, 20, 40, 60\}$ (black, black-dashdot, blue, blue-dashdot) and the open-loop solution to \eqref{eq:OCP} for $N=120$ (red-dashdot). 
        Temperature, emissions, mitigation rate and savings rate. \label{fig:DICE_sim}}
    \end{subfigure}%
    ~ 
    \begin{subfigure}[t]{0.48\textwidth}
        \centering
        \includegraphics[clip, trim= 0.95cm 0.0cm 0.95cm 0.0cm, width=.9375\textwidth]{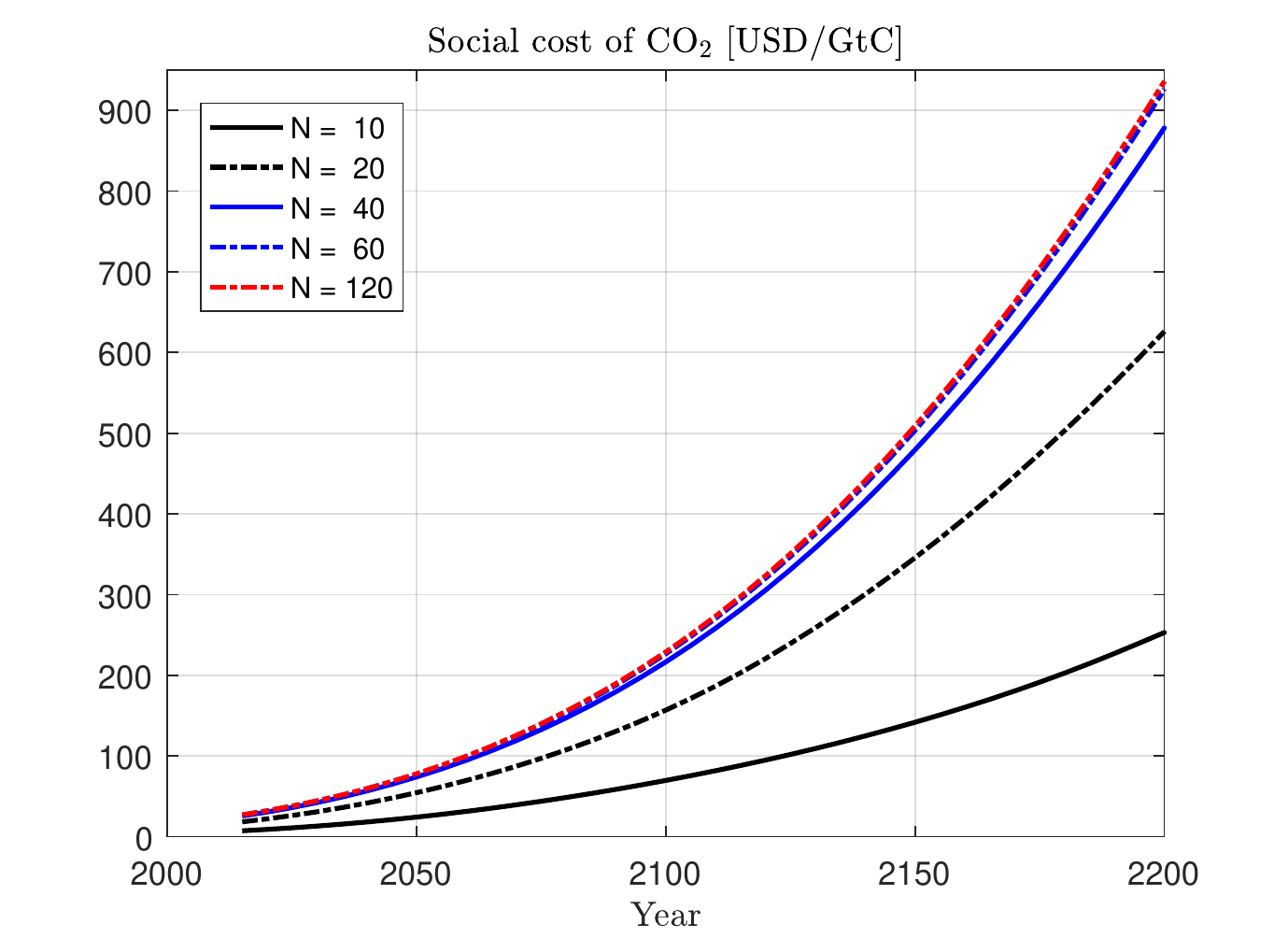}  
        \caption{%Results for \MPCDICE 2013 for varying prediction horizons using Algorithm \ref{algo:MPC-DICE} with $N \in \{10, 20, 40, 60\}$ (black, black-dashdot, blue, blue-dashdot) and the open-loop solution to \eqref{eq:OCP} for $N=120$ (red-dashdot). 
        Social cost of carbon dioxide.\label{fig:SCC_sim}}
    \end{subfigure}
    \caption{Results for \MPCDICE 2016 for varying prediction horizons using Algorithm \ref{algo:MPC-DICE} with $N \in \{10, 20, 40, 60\}$ (black, black-dashdot, blue, blue-dashdot) and the open-loop solution to \eqref{eq:OCP} for $N=120$ (red-dashdot).\label{fig:DICE-MPC}}
\end{figure*}

In order to obtain a receding horizon variant of the original OCP, we define a second optimization problem as follows
\begin{subequations} \label{eq:OCP2}
\begin{align}
\max_{{\bf w}} & \ \  x_{17}(N+1) \\
\textrm{subject to } \nonumber\\
		x(j+1) &= f(x(j), w(j)),  \quad j = 1, \ldots, N  \\
		x(1) &= x^\star(2\,|\,i-1)\\
		w(j) &\in [0,\,1]\times[0,1],\quad j = 1, \ldots, N,		  		  
\end{align}
\end{subequations}
which differs from OCP \eqref{eq:OCP1} in that the initial condition $x(1)$ is available from the previous optimization via the variable $x^\star(2\,|\,i-1)$.\footnote{Here, \textcolor{black}{whenever helpful,} we employ the common MPC notation convention that $x^\star(2\,|\,i-1)$ refers to the second element of the state prediction computed at time $i-1$. } This OCP is to be solved for $i=2, \dots, N_{sim}$, where $N_{sim}$ is the desired simulation horizon.  Consequently, the extra decision variable $v$ is not required, since  $x_{13}(1)=E^\star(2\,|\,i-1)$ and $x_{14}(1) = C^\star(2\,|\,i-1)$. 
% on the decision variables at the same time; i.e., $\mu(1)$ and $s(1)$. 

Solving either OCP \eqref{eq:OCP1} or OCP \eqref{eq:OCP2}, we obtain the following data:
\begin{itemize}
\item The optimal state trajectory $x^\star(j), j = 1, \dots, N+1$, which contains the savings rate and the mitigation rate as 
\[
\mu^\star(j) = x_{15}^\star(j) \text{ \quad and \quad} s^\star(j) = x_{16}^\star(j).
\]
\item The optimal adjoint variables $\lambda^\star_C(j)$ and $\lambda^\star_E(j)$ which are given by the Lagrange multipliers associated to the equality constraints implied by the dynamics of $E(j) = x_{13}(j) $ and $C(j) = x_{14}(j)$.\footnote{The Lagrange multipliers are typically provided by modern NLP solvers such as IPOPT \cite{Waechter06a}, which is used \textcolor{black}{in the open-source DICE implementation  \cite{Faulwasser-MPC-DICE-IAMES2018}.}}
\end{itemize}
Hence, the \SCC \ at time $j$ is obtained by
\[
{\rm SC \mhyphen CO_2}(j) = - 1000 \cdot \frac{\partial W / \partial E(j)}{\partial W / \partial C(j)} = -1000\cdot\frac{\lambda^\star_E(j)}{\lambda^\star_C(j)}.
\]
Finally,  the receding-horizon approximation of  \eqref{eq:OCP} is summarized in Algorithm \ref{algo:MPC-DICE}.

\begin{algorithm}[t]
\caption{MPC-DICE}\label{algo:MPC-DICE}
\begin{algorithmic}[1]
\State Input: simulation horizon $N_{sim}$, prediction horizon $N$,
 \If {$i == 1$} \\
 Solve   OCP \eqref{eq:OCP1}\\
Set $x(1) = x^\star(1|1), \lambda_E(1) = \lambda^\star_E(1|1),  \lambda_C(1) = \lambda^\star_C(1|1)$.
\EndIf
\For {$i  = 2, \dots, N_{sim}$}\\
Solve   OCP \eqref{eq:OCP2} for $x(1) = x^\star(1\,|\,i-1)$. \\
Set $x(i) = x^\star(2\,|\,i-1), \lambda_E(i) = \lambda^\star_E(2\,|\,i-1),  \lambda_C(i) = \lambda^\star_C(1\,|\,i-1)$.
\EndFor
\State Return $x(j)$, $\lambda_E(j)$ and $\lambda_C(j),  j = 0,\dots, N_{sim}$.
\end{algorithmic}
\end{algorithm}

\begin{remark}[Open source code \MPCDICE  \cite{Faulwasser-MPC-DICE-IAMES2018}]
 \textit{MPC-DICE} is an open-source Matlab implementation of \DICE which provides parameter sets for both DICE2013 and DICE2016. Specifically,  \textit{MPC-DICE} provides an implementation of the receding horizon reformulation described above. It uses CasADi \cite{Andersson12a}, which comes with IPOPT \cite{Waechter06a} as an NLP solver, to solve \eqref{eq:OCP}. The relatively simple CasADi syntax enables extensions of the \DICE OCP, some of which we will describe in Section \ref{sec:Constraints}.
 The code is available at \cite{MPC-DICE-Code}.
\end{remark}

Figure \ref{fig:DICE-MPC} shows simulation results obtained with \MPCDICE for the 2016 parameter set for different prediction horizons $N \in \{10, 20, 40, 60\}$ and an \MPCDICE  simulation horizon $N_{sim} = 40$ in comparison to the solution of \eqref{eq:OCP} with $N=120$ of which we plot the first 40 steps. Figure \ref{fig:DICE_sim} shows temperature increase, emissions as well as mitigation rate and savings rate. Figure \ref{fig:SCC_sim} shows the corresponding \SCC trajectories. As one can see, for increasing prediction horizons the receding-horizon input and state trajectories both converge \CMK{towards the infinite-horizon solution; approximated here by computing a long horizon solution ($N =120$).} 
This approximation property can also be observed in Figure \ref{fig:SCC_sim}. Hence we conjecture that under suitable assumptions the approximation properties of MPC, which are established for time-invariant \CMK{and undiscounted} OCPs in \cite{Grun16-DMV}, also hold for \CMK{time-varying and} discounted problems. 

In fact, the theoretical results supporting this conjecture are reasonably mature
where \cite{GrueneNMPCforecon} shows that 
receding horizon control yields approximate optimal solutions for 
discounted problems if the turnpike property holds. Furthermore, it follows 
from a combination of \cite{GGHKW18-Aut} and \cite{GMKW18} that strict dissipativity implies 
the turnpike property for discounted problems, provided the 
discount factor is sufficiently close to one.  Note that the default DICE discount rate of 1.5\%
corresponds to a discount factor of approximately 0.985.  Hence, while these results
for discounted optimal control do not yet accommodate time-varying systems or cost
functions, the primary difficulty lies in checking the appropriate assumptions for 
complicated models such as the DICE model.

\begin{figure*}
    \centering
    \begin{subfigure}[t]{0.48\textwidth}
     %   \centering
        \includegraphics[clip, trim= 0.95cm 0.0cm 0.95cm 0.0cm, width=0.9375\textwidth]{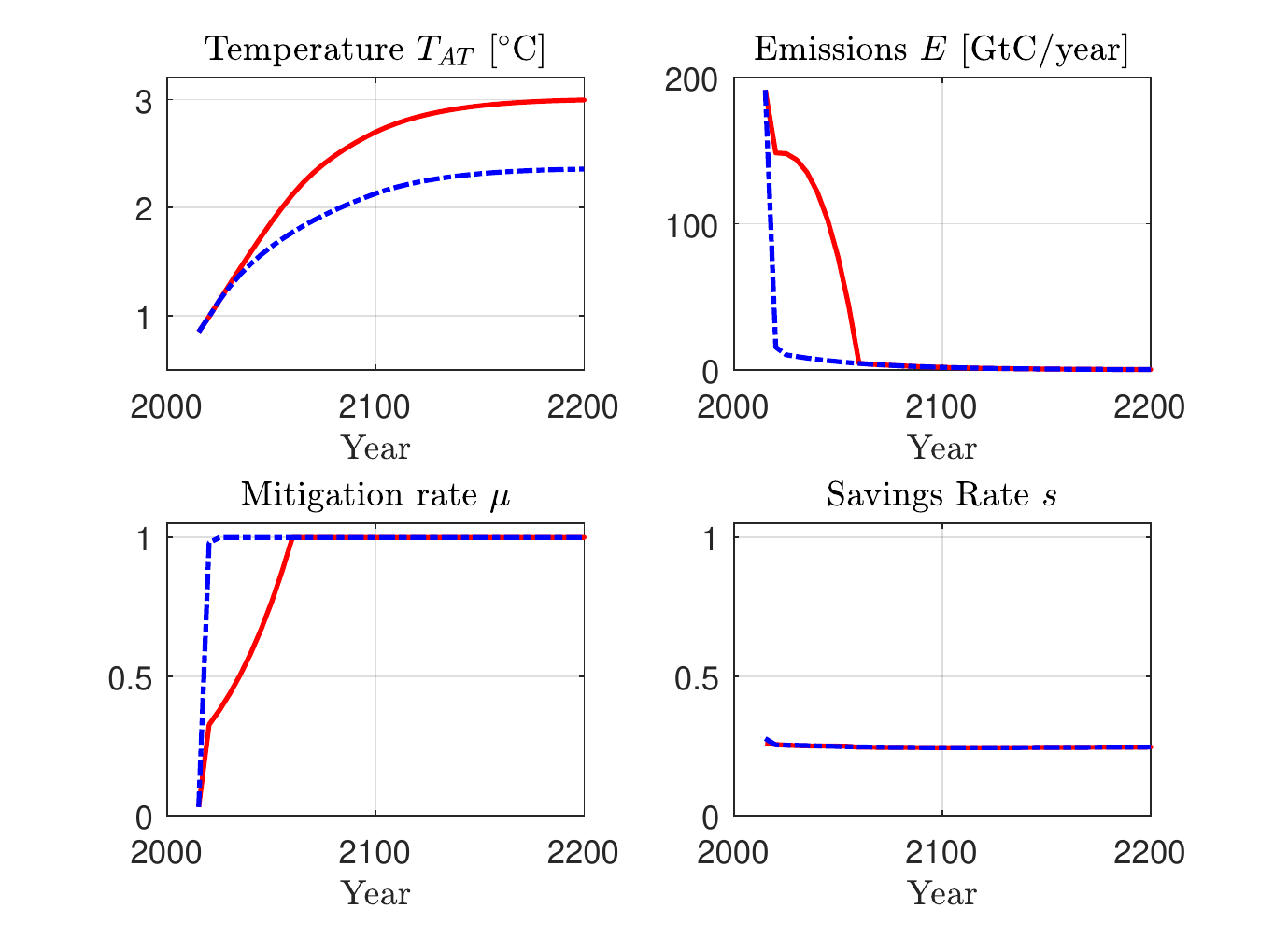} 
        \caption{Results using Algorithm \ref{algo:MPC-DICE} and the
temperature increase constraint \eqref{eq:con_mu_T}  with $2.36\,^\circ$C (blue) and $3\,^\circ$C (red).
\label{fig:T_AT_max_sim}}
    \end{subfigure}%
    ~ 
    \begin{subfigure}[t]{0.48\textwidth}
        \centering
        \includegraphics[clip, trim= 0.95cm 0.0cm 0.95cm 0.0cm, width=0.9375\textwidth]{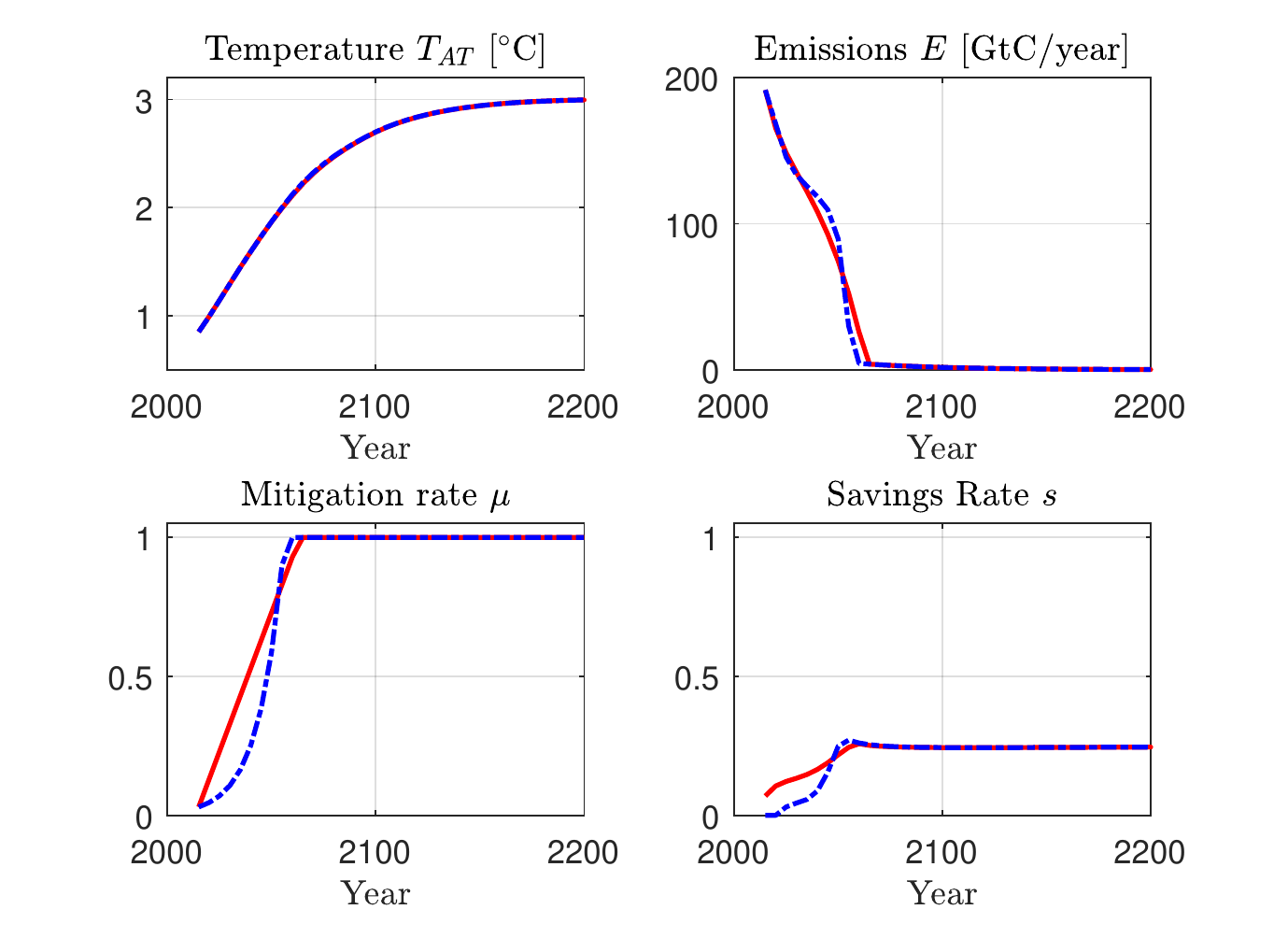}   
        \caption{Results using Algorithm \ref{algo:MPC-DICE}, the  temperature increase constraint \eqref{eq:con_mu_T}  with $3\,^\circ$C, the rate constraint \eqref{eq:con_mu_rate} (red) and the growth constraint \eqref{eq:con_mu_growth} (blue-dashdot).
\label{fig:mu_sim}}
    \end{subfigure}
    \caption{Results for \MPCDICE  2016 with additional constraints. \label{fig:DICE-MPC_con}}
\end{figure*}

\section{State and Input Rate Constraints}
\label{sec:Constraints}

The welfare maximization problem as posed in Section \ref{sec:WM} considers only input magnitude constraints and the dynamics. However, in view of the reports of the IPCC, the overwhelming scientific consensus is that 
temperature increase should be limited to $2\,^\circ$C \cite{AR5SPM} and preferably to $1.5 ^\circ$C \cite{IPCC-1_5-Report}. Inspection of Figure \ref{fig:DICE_sim} reveals that straightforward maximization of social welfare might lead to much higher values of temperature increase in the order of $3-4\,^\circ$C.\footnote{In climate physics the temperature increase is also referred to as the \textit{temperature anomaly}.} 

\CMK{This indicates that there is an inconsistency between the model (or the chosen parameters) and the scientific consensus that $2\,^\circ$C of warming represents a dangerous threshold.  One approach to addressing this is to modify the model directly; for example by changing the climate damages function \eqref{eq:Damages} to reflect the consensus that damages at $2\,^\circ$C are expected to be significantly higher than a loss of 0.9\% of global economic output.  It is also possible to consider a different welfare function that not only places a value on consumption but also values environmental ``services'' (such as clean air and water) \cite{Sterner-Persson-2008} or accounts for the cost of adaptation to climate change \cite{Atolia-etal-IMF-WP-2018}.}
  
\CMK{A third approach, as done in \cite[pp.~69--73]{Nordhaus-2008}, is to add a constraint on the temperature increase to \eqref{eq:OCP}.  As mentioned in \cite[pp.~69--73]{Nordhaus-2008}, the purely economic case for imposing a hard limit is somewhat unjustified as it effectively implies an infinite cost of exceeding the constraint.  However, also as discussed in \cite[pp.~69--73]{Nordhaus-2008}, a hard constraint can represent a tipping point where the climate damages dramatically increase, for example due to adverse climatic effects that are not captured in the simple DICE climate model.  Here, following \cite{WellerCCA2015}, we place an upper limit on the atmospheric temperature rise and investigate what this then requires of the control inputs.}

Consider the state constraint
\begin{subequations} \label{eq:con_mu_T}
\begin{equation} \label{eq:T_AT_max_con}
T_{AT}(i) \leq T_{AT, max}, \quad \forall i \in \mathbb{N}.
\end{equation}
 Moreover, in the economics literature the value for the mitigation rate at time $i=1$ is usually fixed, with
\begin{equation} \label{eq:mu_con}
\mu(1) = \mu_0 = 0.03,
\end{equation} 
\end{subequations}
an estimate of the global greenhouse gas emissions abatement or mitigation rate in the base
year of 2015 (or $\mu(1) = 0.039$ for the DICE2013 base year of 2010) \cite{DICECode}.

Figure \ref{fig:T_AT_max_sim} shows the corresponding results for \MPCDICE 2016 considering \eqref{eq:con_mu_T} and 
\[
T_{AT, max} \in \{2.36\,^\circ C, 3\,^\circ C\}.
\]
 Not surprisingly the tighter temperature target of $2.36\,^\circ$C requires a drastic and fast reduction of emissions, which would imply a steep increase of the mitigation rate in the near future. Interestingly the savings rate is not affected by the temperature target. Moreover, we remark that $T_{AT, max} =2.36\,^\circ C$  is the lowest heuristically determined value of the temperature constraint for which \eqref{eq:OCP} is feasible.

However, the steep increase of the mitigation rate shown in Figure \ref{fig:T_AT_max_sim} might be difficult to realize on a policy level. This motivates analyzing whether constraints on the increase of the mitigation rate are compatible with the temperature target of $2-3\,^\circ$C. To this end, in \cite{WellerCCA2015} (for DICE2013) we considered both rate and growth constraints on $\mu$. The rate constraint takes the form
\begin{subequations}\label{eq:con_rate_growth}
\begin{equation} \label{eq:con_mu_rate}
|\mu(i+1) - \mu(i)| \leq \Delta_\mu, \quad \forall i \in \mathbb{N}
\end{equation} 
while the growth constraint is given by
\begin{equation} \label{eq:con_mu_growth}
\frac{\mu(i+1) - \mu(i)}{\mu(i)} \leq \Gamma_\mu, \quad \forall i \in \mathbb{N}. 
\end{equation}
\end{subequations}

Note that the growth constraint \eqref{eq:con_mu_growth} is motivated by the argument that 
abatement technologies and markets will increase year-on-year rather than in equal increments.

Results for this setting using \MPCDICE with $N_{sim}=40$ and $N = 60$ are depicted in Figure \ref{fig:mu_sim}. We consider the temperature target of $3\,^\circ$C according to \eqref{eq:con_mu_T},  the rate bound $\Delta_\mu = 0.1$ and the growth bound $\Gamma_\mu = 0.53$.
%\footnote{These values have been chosen by try-and-error as for lower ones the temperature target appears to be infeasible.} 
Note that \eqref{eq:mu_con} is needed to ensure that the constraints \eqref{eq:con_rate_growth} are defined at $i=1$.

As one can see the temperature trajectories do not differ much for the different constraints \eqref{eq:con_rate_growth}. However, not surprisingly, the mitigation rates behave very differently as do the emissions. 

It is worth noting that $\Gamma_\mu = 0.53$ is an experimentally determined threshold whereby lower values render
the optimal control problem infeasible.  In this extreme case, the optimal solution shows the savings rate effectively 
taking a value of zero for several years.  Conceptually, this corresponds to an extended period of non-investment in
the capital stock $K$ so as to reduce emissions by explicitly reducing economic activity.

\begin{remark}[Feasibility of temperature targets]
	Given that it is considered feasible that global atmospheric temperature rise could yet be limited
	to $1.5~^\circ C$, it is interesting to note that the default parameters of DICE2016 do not even yield
	a feasible solution to \eqref{eq:OCP} for the higher $2~^\circ C$ limit.  Indeed, in \cite{Nordhaus17-PNAS}
	Nordhaus notes that while the $2~^\circ C$ limit was achievable in DICE2013, this is not the
	case for DICE2016.  However, only limited information on the calibration of the model 
	(i.e., the parameter choices shown in Appendix B) has been provided.  
	In Section~\ref{sec:Opportunities} below,
	we indicate some recent work on improving the transparency of the temperature model 
	parameters.  Similar work on the carbon cycle parameters would be a valuable contribution
	to improved estimates of the \SCC.
\end{remark}

\section{Systems and Control in Climate--Economy Assessment:~Opportunities and Progress}
%\section{Simplified Climate Models and their Applications}
\label{sec:Opportunities}

%At this point is fair to ask how the systems and control community can contribute to  climate--economy assessment. Subsequently we give  a brief account of promising research directions. 

The DICE model interconnects geophysical and socio\-economic dynamics, the structure and parameters of which are clearly subject to a vast array of uncertainties. Quantifying the implications of these uncertainties on \SCC estimates and on related policy advice is consequently an issue of major importance to policymakers. Given the wide range of tools developed by the systems and control community for handling and quantifying uncertainty, many challenges and research opportunities exist for this community within climate--economy assessment. 
In this section, we identify several key avenues of research opportunity, and describe previous work undertaken in the context of those directions.

\subsection{Uncertainty Quantification}
In early 2017, the U.S.\ National Academies released an extensive and influential report on improving estimates of the \SCC \cite{NatAcadsUpdatingSCC}.  Many of the recurring themes in \cite{NatAcadsUpdatingSCC} would be familiar to the systems and control community, particularly around the quantification of uncertainty and its impacts.

The climate economics literature distinguishes between \emph{parametric} uncertainty \cite{Anderson-2014,Gruene-2010,Botzen-2012,InteragencyWorkingGroup} and \emph{structural} uncertainty \cite{Gillingham2015,Moore-2015,Guivarch-2018,Nordhaus17-PNAS} in IAMs such as DICE. Parametric uncertainty is uncertainty about the value of various parameters within an IAM module, e.g.\ $F_{2\times}$ or $\Phi_T$ in \eqref{eq:Climate}. Structural uncertainty, on the other hand, refers to uncertainty regarding the functional form of the equations comprising the IAM. As one example of structural uncertainty, consider carbon cycle feedbacks---currently neglected in DICE---wherein rising surface temperatures lead to thawing of carbon-rich permafrost and the consequent release of methane, itself a potent greenhouse gas.
 
 \subsection{Identification of Predictive Climate Models}
The geophysical models presented in Section \ref{sec:dynamics} are
clearly significant simplifications of reality, with the climate (temperature)
model having been originally proposed in \cite{Schneider1981Geophysic}. 
While low-order models are necessary to efficiently solve the optimal control
problem \eqref{eq:OCP}, it is possible to construct improved higher order 
models. In particular, a large number of supercomputer-based, atmosphere--ocean general circulation models (AOGCMs) have been developed by a number of climate modeling centres, providing very high spatio-temporal resolution.  Furthermore, many of these AOGCMs participate in the Coupled Model Intercomparison Project (CMIP) \cite{Taylor2012}, which effectively provides input--output data for a number of AOGCMs.

With such input--output data available, standard system identification tools
can be applied.  In \cite{WellerIFAC2014}, for example, we derived fourth-order linear 
time-invariant models from the CMIP3 (CMIP, Phase 3) data set. In particular, 12 AOGCMs from the CMIP3 ensemble were identified for which linear, time-invariant (LTI) models of order 4 were able to very closely approximate surface temperature projections under each of the four Representative Concentration Pathway (RCP) emission scenarios in the AR5 assessment report (see \cite[p.\ 45, Box SPM.1]{AR5SPM}).

The LTI models identified in \cite{WellerIFAC2014} are suitable for application in feedback-based approaches to mitigation; see for example \cite{WellerAUCC2014} in which these models are applied in an optimal control-based approach to geoengineering of the climate based on solar radiation management (SRM). In \cite{WellerAAEC2015}, we considered \eqref{eq:Climate}--\eqref{eq:Capital} in which the climate model \eqref{eq:Climate} is replaced by each of 12 fourth-order models derived in \cite{WellerIFAC2014}. The range of estimated \SCC values for 2015 obtained using this method span US\$10.20--\$58.20/tCO2 depending on the specific CMIP3 model, with an ensemble mean \SCC of US\$22.90/tCO2. This wide range of values highlights the substantial variability in estimates of the \SCC arising from scientific uncertainty in the climatic response to net radiative forcing, all other components of the IAM being held constant.

In recognizing the numerous uncertainties inherent in estimation of the \SCC, the National Academies report recommends that research effort on the \SCC be focused on ``incorporating the most important sources of uncertainty, rather than seeking to incorporate all possible sources of uncertainty'' \cite[p.~54]{NatAcadsUpdatingSCC}. One approach along these lines avoids direct appeal to strongly geophysically-inspired climate models, instead capturing climate behaviour simply via three scalar parameters: the transient climate response (TCR), the equilibrium
climate sensitivity (ECS), and $F_{2\times}$.

Here we recall ECS as the steady-state atmospheric temperature arising from a doubling of atmospheric carbon, $F_{2\times}$ as the associated downward radiative forcing at top-of-atmosphere for doubled atmospheric carbon (see Remark~\ref{rem:lambda}), and define TCR as the temperature change at the time of CO$_2$ doubling under a scenario in which CO$_2$ concentrations increase by $1$\%~$\mathrm{yr}^{-1}$.

In \cite{HafeezAuCC2016} we proposed an optimization-based methodology for computing the parameters of a climate model in such a way that the resulting model exhibits a specified TCR. The results reported in \cite{HafeezAuCC2016} enable policymakers using DICE---which specifies the TCR parameter only indirectly---to compute optimal CO$_2$ emissions pathways which directly reflect the reported TCR of state-of-the-art AOGCM climate models documented in the most recent (Fifth) Assessment Report (AR5) of the IPCC (see 
\cite[p.\ 818, Table 9.5]{AR5ClimateModels}).

\subsection{Modular Tools for Simulation and Optimization}

The report \cite{NatAcadsUpdatingSCC} also recommends increasing transparency around the models used, and maintaining modular models to allow for advances in any particular model to be easily incorporated.  As an example of this latter topic, in \cite{Faulwasser-FAIR-DICE-IAMES2018}
we replaced the standard DICE geophysical model (i.e., \eqref{eq:Climate}--\eqref{eq:Carbon}) with a state-of-the-art reduced-order geophysical model termed FAIR \cite{Millar17}.  In this context it is worth noting that the original reference for the FAIR model  \cite{Millar17} does not highlight the fact that FAIR is a system of differential algebraic equations (DAEs), which should be accounted for in developing simulation code.  
%At large much more effort on the development of modular tools is needed.% in this direction.
%And arguably it is the systems and control community who has the means to do so. 
%catalyzed the development of powerful yet generic  tools for simulation and optimization; many of which are open source\cite{}. 

However, as of now, when it comes to uncertainty quantification combined with dynamic optimization no widely accepted open-source tools that go beyond various sampling techniques exist. Hence, there is a need for tailoring and implementing the powerful methods developed by the systems and control community to climate-economy assessment.

\subsection{System Theoretic Analysis}

In addition to the research questions mentioned above and posed in \cite{NatAcadsUpdatingSCC}, the framework of 
\textit{discounted} optimal control (i.e., where the cost function involves a discount factor) is
one which has received less attention in the systems and control community than the usual undiscounted framework.  Several recent results \cite{Gaitsgory2015,GGHKW18-Aut,GKW17-JOTA,GMKW18} indicate
that the connections between strict dissipativity, turnpike properties, and numerically
accurate approximations via MPC, which are known for undiscounted optimal control (as reported in
\cite{Grun16-DMV}) also hold in the discounted setting.  However, checking the necessary assumptions
to use results in particular applications, such as for the DICE model, remains a difficult problem.

Moreover, the fact that the receding-horizon solution to DICE approximates long/infinite horizon solutions quite well  (see \cite[Fig.\ 2]{HafeezIFAC2017}) gives raise to the conjecture that the DICE OCP exhibits a time-varying turnpike phenomenon. However, a formal analysis remains to be done. 

Finally, recall that Nordhaus also proposed a regionally distributed variant of DICE named RICE, wherein several economic regions (US, EU, China, ...) are considered \cite{Nordhaus-PNAS2010}. From a systems and control perspective RICE raises many interesting problems ranging from distributed implementation to game-theoretic frameworks.  

\section{Summary and Concluding Remarks}
\label{sec:Summary}
The overwhelming scientific consensus is that avoiding the worst potential effects of
anthropogenic climate change require achieving economy-wide net-zero greenhouse gas emissions
by the middle of this century.  Such a significant economic transition will require a suite of policy
responses, many of which will rely on a price on greenhouse gas emissions \cite{Kellett-Humboldt2018}.  
Estimates of the \SCC provide guidance on the range of prices.

In this paper, we have provided a complete tutorial description of the DICE model, one of the most widely used IAMs
for estimation of the \SCC and have indicated some work already undertaken to improve
\SCC estimates and indicated where we believe the systems and control community can make
important contributions.

%Specifically, we have used MPC both as a numerical solution technique and
%as a method to cope with the inherent uncertainty in the DICE submodels.  

%FAIR-DICE \cite{Faulwasser-FAIR-DICE-IAMES2018}, uncertainty quantification, discounted optimal control

\section*{Appendix A: Default initial conditions}
\begin{center}
\begin{tabular}{|c||c|c|c|} \hline
 		& $T_{\rm AT}(0)$ & $T_{\rm LO}(0)$ & $K(0)$ \\ \hline\hline
2013R	& 0.8	 & 0.0068	 & 135  \\ \hline
2016R	& 0.85 & 0.0068  & 223 \\ \hline
\end{tabular}
\end{center}

\begin{center}
\begin{tabular}{|c||c|c|c|} \hline
 		& $M_{\rm AT}(0)$ & $M_{\rm UP}(0)$ & $M_{\rm LO}(0)$  \\ \hline\hline
2013R	& 830.4 & 1527 & 10010   \\ \hline
2016R	& 851 & 460 & 1740  \\ \hline
\end{tabular}
\end{center}

The parameters for calculating $\sigma_0 = \frac{e_0}{q_0(1-\mu_0)}$:
\begin{center}
\begin{tabular}{|c||c|c|c|} \hline
 		& $e_0$		& $q_0$		& $\mu_0$ \\ \hline\hline
2013R	& 33.61	& 63.69	& 0.039 \\ \hline
2016R	& 35.85	& 105.5	& 0.03 \\ \hline
\end{tabular}
\end{center}

\section*{Appendix B: Default parameter values}
\begin{table*}
\renewcommand{\arraystretch}{1}
\begin{center}
\caption{Default parameter values for DICE 2013 and DICE 2016.}
\begin{tabular}{|c||c|c| c| c |} \hline
		Parameter & Value               	& Value		& Equations & Unit \\
				 &   DICE2013R   	& DICE2016R	& & \\ \hline \hline
%				 &				&			 \\ \hline
		$\Delta$	& 5		& 5 & \eqref{eq:time_step} & years \\ \hline
		$t_0$	& 2010	& 2015 & \eqref{eq:time_step} & year \\ \hline
		$N$		& 60		& 100 &\eqref{eq:OCP1}, \eqref{eq:OCP2} & time steps	  \\ \hline
		$\mu_0$ 	& 0.039	& 0.03	& \eqref{eq:mu_con} & \\ \hline \hline
	\multicolumn{5}{|l|}{Climate diffusion parameters} \\ \hline\hline
		$\phi_{11}$ & 0.8630 & 0.8718 & \eqref{eq:Climate}, \eqref{eq:Climate_params} &  \\
		$\phi_{12}$ & 0.0086 & 0.0088  &\eqref{eq:Climate}, \eqref{eq:Climate_params}&\\
		$\phi_{21}$ & 0.025 & 0.025 &\eqref{eq:Climate}, \eqref{eq:Climate_params}& \\
		$\phi_{22}$ & 0.975 & 0.975  &\eqref{eq:Climate}, \eqref{eq:Climate_params}&\\ \hline \hline
	\multicolumn{5}{|l|}{Carbon cycle diffusion parameters} \\ \hline \hline
		$\zeta_{11}$ & 0.912 & 0.88  &\eqref{eq:Carbon}, \eqref{eq:Carbon_params} &\\
		$\zeta_{12}$ & 0.03833 & 0.196 &\eqref{eq:Carbon}, \eqref{eq:Carbon_params} &\\
		$\zeta_{21}$ & 0.088 & 0.12  &\eqref{eq:Carbon}, \eqref{eq:Carbon_params} &\\
		$\zeta_{22}$ & 0.9592 & 0.797  &\eqref{eq:Carbon}, \eqref{eq:Carbon_params} &\\
		$\zeta_{23}$ & 0.0003375 & 0.001465 &\eqref{eq:Carbon}, \eqref{eq:Carbon_params} & \\
		$\zeta_{32}$ & 0.00250 & 0.007  &\eqref{eq:Carbon}, \eqref{eq:Carbon_params} &\\
		$\zeta_{33}$ & 0.9996625 & 0.99853488&\eqref{eq:Carbon}, \eqref{eq:Carbon_params} &  \\ \hline \hline	
	\multicolumn{5}{|l|}{Other geophysical parameters} \\ \hline \hline
		$\eta$ & 3.8 & 3.6813  & \eqref{eq:Climate}, Remark~\ref{rem:lambda} & W/m$^{2}$\\
		$\xi_1$ & 0.098 & 0.1005  & \eqref{eq:Climate}, \eqref{eq:xi1}  &\\
		$\xi_2$ & 12/44 & 12/44 &\eqref{eq:Carbon}, \eqref{eq:Carbon_params}& GtC/GtCO$_2$ \\
		$M_{\rm AT, 1750}$  & 588 & 588 & \eqref{eq:Climate}, Remark~\ref{rem:lambda} & GtC \\ \hline
		$f_0$	& 0.25	& 0.5  & \eqref{eq:FEX} & W/m$^2$ \\ 
		$f_1$	& 0.70	& 1.0  & \eqref{eq:FEX} & W/m$^2$ \\ 
		$t_f$		& 18		& 17  & \eqref{eq:FEX} & time steps \\ \hline
		$E_{L0}$	& 3.3		& 2.6  &\eqref{eq:eland} & GtCO$_2$/yr \\ 
		$\delta_{EL}$ & 0.2	& 0.115  & \eqref{eq:eland} & \\ \hline \hline
	\multicolumn{5}{|l|}{Socioeconomic parameters} \\ \hline \hline
		$\gamma$ & 0.3 & 0.3 & \eqref{eq:Capital}, \eqref{eq:output} &\\
		$\theta_2$ & 2.8 & 2.6  & \eqref{eq:Capital}, \eqref{eq:Net_Economic_Output} & n/a \\
		$a_2$	& 0.00267 & 0.00236 & \eqref{eq:Capital}, \eqref{eq:Net_Economic_Output}&\\
		$a_3$	& 2		& 2 & \eqref{eq:Capital}, \eqref{eq:Net_Economic_Output}& n/a \\
		$\delta_K$ & 0.1 & 0.1 & \eqref{eq:Capital}, \eqref{eq:PhiK}&\\ \hline
		$\alpha$ & 1.45 &1.45 & \eqref{eq:Utility} &\\
		$\rho$ & 0.015 & 0.015 & \eqref{eq:OCP} &\\ \hline
		$L_0$	& 6838	& 7403  & \eqref{eq:Population} & millions people \\ 
		$L_a$	& 10500	& 11500  &\eqref{eq:Population} & millions people \\ 
		$\ell_g$	& 0.134	& 0.134 &\eqref{eq:Population} &\\ \hline
		$A_0$	& 3.80	& 5.115  &\eqref{eq:TFP} &\\ 
		\CMK{$g_A$}	& 0.079	& 0.076  &\eqref{eq:TFP} &\\ 
		$\delta_A$ & 0.006	& 0.005  &\eqref{eq:TFP} &\\ \hline
		$\sigma_0$ & 0.5491 & 0.3503 & \eqref{eq:EI} & GtC / trillions 2010USD \\ 
		$g_\sigma$	& 0.01	& 0.0152  & \eqref{eq:EI} &\\ 
		$\delta_\sigma$ & 0.001	& 0.001 & \eqref{eq:EI} &\\ \hline
		$p_b$	& 344	& 550  & \eqref{eq:theta1} & 2010USD/tCO$_2$\\ 
		$\delta_{pb}$ & 0.025 & 0.025 & \eqref{eq:theta1} &\\ \hline
		$scale1$ & 0.016408662 & 0.030245527  & \eqref{eq:scaling} & n/a \\
		$scale2$ & 3855.106895 & 10993.704  & \eqref{eq:scaling} & 2010USD \\ \hline
\end{tabular}
\end{center}
\end{table*}
Rather than the cost function in \eqref{eq:OCP}, Nordhaus has used the scaled cost 
function
\begin{align}
	\label{eq:scaling}
	  (scale2) + (scale1) \cdot \max_{{\bf s}, {\bf \mu}}&  \sum_{i=1}^{N+1} \frac{U(C(i),L(i))}{(1+\rho)^{\Delta i}}  .
\end{align}
This obviously has no impact on the solution of \eqref{eq:OCP}.  These values are chosen so that the 
optimal value function has a numerical value consistent with economic intuition.

\begin{ack}                               % Place acknowledgements here.
	The authors are supported by the Australian Research Council under
	Discovery Project DP180103026.
	TF acknowledges financial support from the Daimler Benz Foundation.
\end{ack}

\bibliographystyle{plain}        % Include this if you use bibtex 
\bibliography{AnnRevControl-ClimateEconomy}           % and a bib file to produce the 
                                 % bibliography (preferred). The
                                 % correct style is generated by
                                 % Elsevier at the time of printing.

%\appendix
%\section{A summary of Latin grammar}    % Each appendix must have a short title.
%\section{Some Latin vocabulary}         % Sections and subsections are supported  
                                        % in the appendices.
\end{document}